\documentclass[prc,aps,amsmath,amssymb,letter,showpacs]{revtex4}
\usepackage{graphicx}
\usepackage{epsfig,rotate,latexsym}
\usepackage{hyperref}
\usepackage{rotating}
\usepackage{subfig}
\begin{document}

\title{Effect of inverse magnetic catalysis on conserved charge fluctuations 
in hadron resonance gas model}

\author{Ranjita K. Mohapatra}

\email{ranjita@iitb.ac.in}

\affiliation{Department of Physics, Indian Institute of Technology Bombay,
Mumbai 400076, India}%

\affiliation{Institute of Physics, Bhubaneswar 751005, India}

\affiliation {Homi Bhabha National Institute, Training School Complex,  
Anushakti Nagar, Mumbai 400085, India}%

\begin{abstract}

The effect of inverse magnetic catalysis (IMC) has been observed on the conserved 
charge fluctuations and the correlations along the chemical freeze-out curve in a 
hadron resonance gas model.
The fluctuations and the correlations have been compared with and without charge conservations. 
The charge conservation plays an important role in the calculation of the fluctuations
at nonzero magnetic field and for the fluctuations in the strange charge at 
zero magnetic field. The charge conservation diminishes the correlations $\chi_{BS}$ 
and $\chi_{QB}$, but enhances the correlation $\chi_{QS}$.  
The baryonic fluctuations (2nd order) at $B = 0.25$ ${GeV}^2$ increases more than two 
times compared to $B = 0$ at higher $\mu_{B}$. The fluctuations have been compared at 
nonzero magnetic field along the freeze-out curve 
i.e along fitted parameters of the chemical freeze-out temperature and chemical potentials,
with the fluctuations at nonzero magnetic field along the freeze-out curve with the IMC effect, 
and the results are very different with the IMC effect. This is 
clearly seen in the products of different moments ${{\sigma}^2}/{M}$ and 
$S\sigma$ of net-kaon distribution.   

\end{abstract}

\pacs{25.75.-q, 12.38.Mh}

\maketitle

\section{INTRODUCTION}

The ultimate goal of ultra relativistic heavy ion collisions is to study 
the phase structure of a strongly interacting system in QCD at finite 
temperature $T$  and finite baryon chemical potential $\mu_{B}$. Lattice QCD 
shows a crossover from the confined phase to the quark gluon plasma (QGP) phase 
at high temperature and low baryon chemical potential \cite{aoki}.
However, the system exhibits a first order phase transition from the confined phase
to the QGP phase at low temperature and high baryon chemical potential 
\cite{yazaki,ejiri}. It is believed that there exists a QCD critical point 
when the first order transition line ends and this is a very challenging 
task to find the critical point in lattice QCD as well as from the 
experimental results on fluctuations of conserved quantities.

The basic features of the physical system created at the time of 
chemical freeze-out in heavy ion collisions are well described in terms of 
the hadron resonance gas (HRG) model \cite{munzi1,munzi2}. 
There is an excellent agreement between experimental data on particle ratios
in heavy ion collisions with corresponding thermal abundances
calculated in the HRG model at appropriately chosen temperature
and baryon chemical potential with different conserved charges taken 
into account \cite{munzi3}. The universal chemical freeze-out curve in the $T-\mu_{B}$ plane 
is determined by the condition $E/N=\epsilon/n\simeq$ 1 GeV, where $E (\epsilon)$ is the 
internal energy (density) and $N (n)$ is the particle number (density) \cite{cleymans}.
It has been already proposed that event-by-event fluctuations of conserved quantities 
such as net baryon number, net electric charge, and net strangeness
as a possible signal of the QGP formation and quark-hadron 
phase transition \cite{asakawa,koch}. The fluctuations in net electric charge are 
suppressed in the QGP phase compared to the hadronic gas phase due to the fractional 
charge carriers in the QGP phase compared to unit charge carriers in the  
hadronic gas phase. The fluctuations originated in the QGP phase may survive
until the freeze-out due to the rapid expansion of the fireball and can be exploited
as a signal of the QGP formation in the early stages of the relativistic heavy ion 
collisions \cite{asakawa,koch}.

Moreover, higher order moments of conserved charge fluctuations are more
sensitive to the large correlation lengths in the QGP phase and relax slowly to their 
equilibrium values at the freeze-out \cite{stephanov}. The divergence of correlation
length or higher order fluctuations will hint towards the existence of a critical 
point in QCD phase diagram. So, higher moments, different
ratios and skewness and kurtosis of conserved charges have been measured experimentally
and compared with the HRG model predictions along the freeze-out curve \cite{redlich}.        
The deviation of experimental results from the HRG model predictions may conclude the
presence of non hadronic constituents or non thermal physics in the primordial medium
\cite{alice1}. Ratios of susceptibilities in lattice QCD have been shown to be 
consistent with the HRG model predictions near zero chemical potential \cite{gavai,cheng}.

However, it is very important to study these fluctuations in the presence of magnetic field
because of the huge magnetic field produced in non central relativistic heavy ions
collision due to the valence charges of colliding 
nuclei \cite{skokov}. The study of chiral magnetic effect (CME) also has 
drawn a lot of attention in heavy ion community due to the presence of the huge 
magnetic field in the QGP phase \cite{fuku3}. One would expect electric charge separation 
with respect to the reaction plane due to CME. This magnetic field decreases with the time given by
$eB(t)=eB_{0}[1+(t/t_{0})^2]^{-3/2}$, where $eB_{0}$ is the maximum magnetic field
produced and varies as $(0.05 GeV)^2 (1fm/b)^2 Z \sinh Y$ \cite{kharzeev,fuku}. 
Here b is the impact parameter, Z is the atomic number of nuclei and the beam 
rapidity is approximated as $\sinh Y\simeq \sqrt{s_{NN}}/{(2m_{N})}$. As the center 
of mass collision energy increases, the initial value of the magnetic field increases
and becomes $\simeq 10^{20}$ Gauss for LHC energies. The life time parameter $t_0$
decreases with increasing energy as $t_{0}\simeq {b/(2\sinh Y)}$. So, the magnetic field
decreases with time as $t^{-2}$ and the effect of this magnetic field may not play 
an important role on the fluctuation of conserved charges along the freeze-out curve.
However, it has been shown that magnetic fields of similar magnitude, as the maximum 
value of the field can arise from induced currents due to rapidly decreasing 
external field and may sustain for longer time \cite{tuchin}. The effect of magnetic 
field on the conserved charge fluctuations have been studied in the HRG model along 
the universal freeze-out curve and compared with the available experimental 
data \cite{rajarshi}.     

However, it has been shown that the IMC effect arises in 
the presence of an external magnetic field in lattice QCD, in which the chiral 
transition temperature decreases \cite{bali1}. But the system exhibits 
magnetic catalysis at zero temperature where the chiral condensate 
increases in an external magnetic field. The IMC effect might be due to the decrease
in interaction strength in the presence of the magnetic field \cite{gatto}. This decrease
of interaction strength is consistent with asymptotic freedom of QCD if the relevant
scale $\sqrt{eB} \simeq\Lambda_{QCD}$ \cite{mian,pawl}. Since the chiral transition 
temperature decreases in the presence of the magnetic field, the freeze-out curve in the
$T-\mu_{B}$ plane will correspond to a lower temperature \cite{fuku2} in the HRG 
model. It has been shown that electric charge conservation and strangeness conservation 
play an important role at higher baryon chemical potential in nonzero magnetic field. 
Electric charge susceptibility along the freeze-out curve is very large without 
charge conservation at higher $\mu_{B}$, but it decreases significantly when the charge 
conservation is taken into account \cite{fuku2}. So, it is very important to consider electric 
charge conservation and strangeness conservation at higher $\mu_{B}$ in 
the presence of the magnetic field.

In this work, one considers the IMC effect to lower the freeze-out temperature in HRG model
in the presence of the external magnetic field. Then fluctuations and correlations are measured 
at that temperature with and without charge conservation at nonzero magnetic field.

This paper is organized as follows.
The essential aspects of the HRG model have been discussed in the presence of external 
magnetic field in section II. Section III describes the different universal freeze-out 
conditions. Section IV describes the conserved charge densities and the fluctuations 
along the freeze-out curve. I also discuss beam energy dependence of the products of 
moments along the freeze-out curve and compare it with the available STAR data. Then 
the conclusions have been presented in Section V.  

\section{HRG MODEL IN THE PRESENCE OF THE MAGNETIC FIELD}

The HRG model in the presence of the magnetic field has been studied and the thermodynamic 
quantities like the pressure, the energy density, the entropy density, the magnetization and 
the speed of sound are presented as functions of the temperature and the magnetic field
\cite{endrodi}. The basic quantity in the HRG model is the grand partition function
defined for each hadron species i as 

\begin{equation}
{\ln Z_{i}}=\pm {Vg_{i}\int{\frac{d^3p}{(2\pi)^3}\ln\left[1\pm{e^{-(E_{i}-\mu_{i})/T}}\right]}} 
\end{equation}

Here $\pm$ corresponds to fermions and bosons respectively. Here V is the volume of 
the system, $g_{i}$ is the spin degeneracy factor, $E_{i}=\sqrt{p^2+{m_{i}}^2}$ is the single 
particle energy and $\mu_{i}=B_{i}\mu_{B}+S_{i}\mu_{S}+Q_{i}\mu_{Q}$ is the 
chemical potential. Here $B_{i}$, $S_{i}$ and $Q_{i}$ are the baryon number, the strange
and the electric charge of the particle and $\mu_{B}$, $\mu_{S}$ and $\mu_{Q}$ are the 
corresponding chemical potentials. The strangeness and the
electric charge chemical potentials have been introduced to implement
the conservation laws of strangeness and electric
charge for the entire system. $\mu_{S}$ and $\mu_{Q}$ have finite value to
obtain total strangeness $N_{S}=0$ and $\frac{B}{Q}\simeq 2.52$. 
I have incorporated all the hadrons listed in the
particle data book upto mass 3 GeV \cite{pdg2012}. All the thermodynamic quantities 
like the pressure, the energy density and the entropy density etc. can be derived from this 
partition function.

For a constant magnetic field along the Z-axis, the well known phenomena of Landau 
quantization of energy levels for a charged particle takes place along the 
plane perpendicular to the magnetic field \cite{landau}. The single particle energy
for a charged particle in the presence of the magnetic field is given by 
$ E=\sqrt{{p_{z}}^2+m^2+2|qB|(n+1/2-s_{z})}$. Here n is the Landau level and $s_{z}$ is the
z component of the spin of the hadron. 
 
The grand partition function in the presence of magnetic field is given by

\begin{equation}
{\ln Z_{i}}=\pm {V \sum_{s_z=-s}^{+s} \sum_{n=0}^{\infty} \frac{|qB|}{2\pi}
\int{\frac{dp_{z}}{2\pi}\ln\left[1\pm{e^{-(E_{i}-\mu_{i})/T}}\right]}} 
\end{equation}

The vacuum part is in general divergent and it needs to be regularised and 
renormilised \cite{endrodi}. Since one is interested in the fluctuations and 
the correlations at the time of freeze-out, one can safely ignore the vacuum part.
 
\section {UNIVERSAL FREEZE-OUT CURVE}

Certain properties of the thermal fireball are same at the chemical freeze-out at all 
collision energies. These common properties provide the unified chemical freeze-out
conditions in heavy ion collisions. The first unified condition for the chemical freeze-out 
is given by  $E/N=\epsilon/n\simeq$ 1 GeV. This makes sense because the freeze-out occurs
when the average energy per particle $\sim m+\frac{3}{2}T$ crosses 1 GeV for the non-relativistic 
particles. The second condition for the unified chemical freeze-out is given by a fixed value 
for the sum of baryon and anti-baryon densities i,e $n_{B}+n_{\bar B}\simeq 0.12$ $fm^{-3}$ 
\cite{stachel} and the third unified chemical freeze-out is given by a fixed value of 
entropy density i.e $s/T^3\simeq 7$ \cite{cleyman2}.
   
The magnetic field produced in the relativistic heavy ion collisions at LHC energies
can reach of the order of 0.25 ${GeV}^2$.
Keeping this in mind, I have shown the effect of the inverse magnetic catalysis on the freeze-out
curve in the HRG model at $B= 0.25$ ${GeV}^2$ in Fig.1.
I have compared the freeze-out curve determined by the condition
$E/N=\epsilon/n\simeq$ 1 GeV for zero and nonzero magnetic field with charge
conservation and without charge conservation in Fig.1a. A similar plot is already shown
in \cite{fuku2}, since this is an important part of the analysis presented here,
Fig.1a has been shown for completeness. The solid line in Fig.1a represents the
freeze-out curve without charge conservation and the dashed line represents the
freeze-out curve with charge conservation. One can clearly see the
freeze-out temperature is lowered at nonzero B due to the effect of
the inverse magnetic catalysis with charge conservation. However, the freeze-out
temperature at nonzero B increases at higher $\mu_{B}$ without charge conservation.
This is due to the fact that at higher $\mu_{B}$ there are more baryons, particularly
more protons at nonzero B. If there is no charge conservation, then number density increases
due to more protons produced at nonzero B. So, the freeze-out curve determined by the constant
$E/N\simeq$ 1 GeV should be pushed to higher temperature \cite{fuku2}.

Fig.1b represents the freeze-out curve determined from the condition 
$n_{B}+n_{\bar B}\simeq 0.12$ $fm^{-3}$. The freeze-out temperature is lowered at nonzero 
magnetic field with and without charge conservation. However, the freeze-out temperature
decreases sufficiently at nonzero B without charge conservation at higher $\mu_{B}$. This 
is due to the fact that the proton density increases sufficiently 
at nonzero B and higher $\mu_{B}$ when 
there is no charge conservation. So, to keep $n_{B}+n_{\bar B}\simeq 0.12$ $fm^{-3}$ fixed,
the temperature will be pushed to downwards.

Fig.1c represents the freeze-out curve determined from the condition $s/T^3\simeq 7$.
The freeze-out temperature is also lowered at nonzero B with this freeze-out condition.
However, the freeze-out temperature is very low ($\sim 20$ $MeV$) at nonzero B without 
charge conservation at higher $\mu_{B}=600$ $MeV$. At higher $\mu_{B}$, the entropy
density decreases sufficiently due to the increase in baryon density (proton density)
at nonzero B. To keep the ratio $s/T^3\simeq 7$ fixed, temperature should decrease
sufficiently. The chemical freeze-out parameters determined from
the different universal conditions are almost same at zero B. But, they are very 
different at nonzero B at higher $\mu_{B}$ without
charge conservation. From all these plots, it is clear that it is very important 
to use charge conservation at nonzero B to determine the chemical freeze-out parameters from
different freeze-out condition.     
   
\begin{figure}
\centering
\subfloat{\includegraphics[scale=0.45]{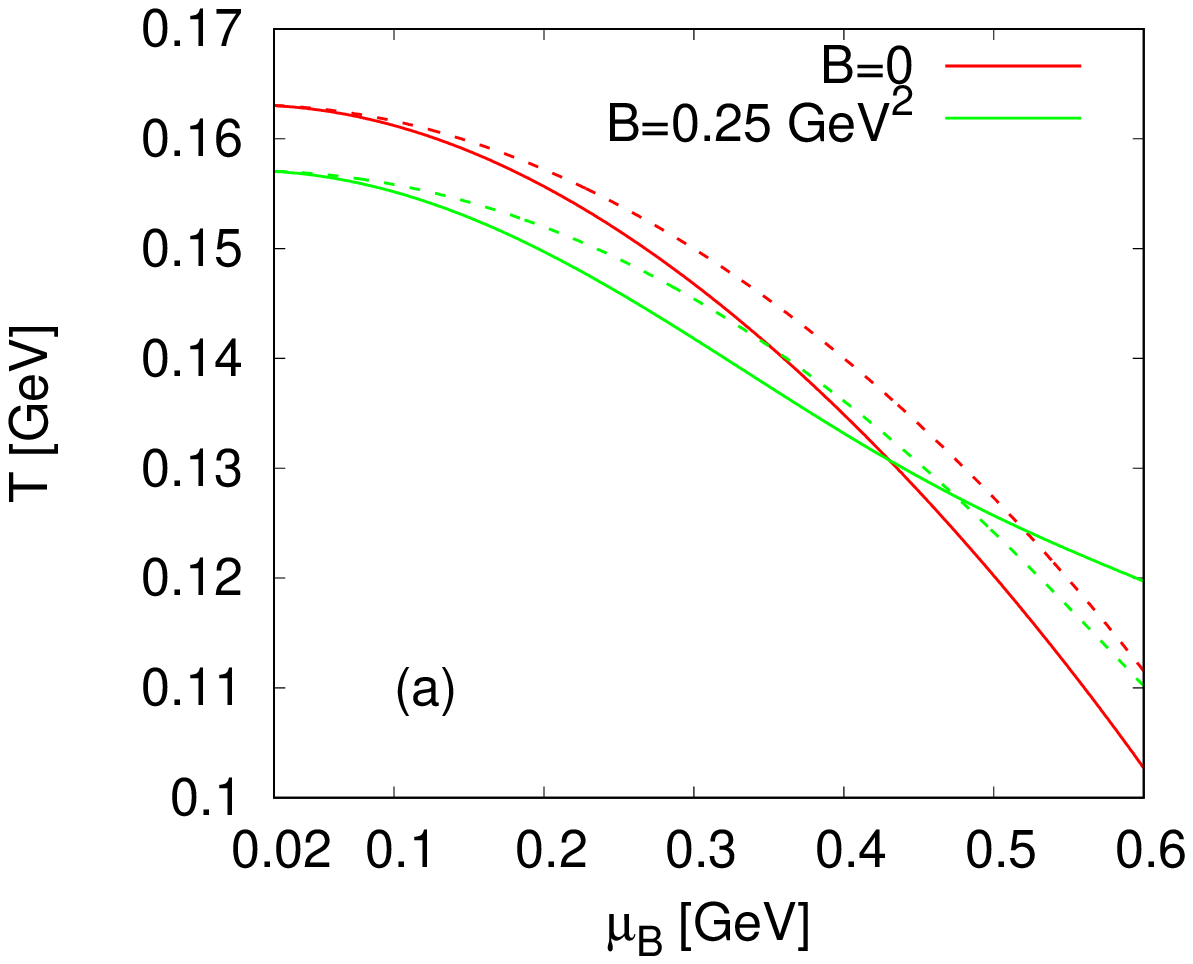}}
\subfloat{\includegraphics[scale=0.45]{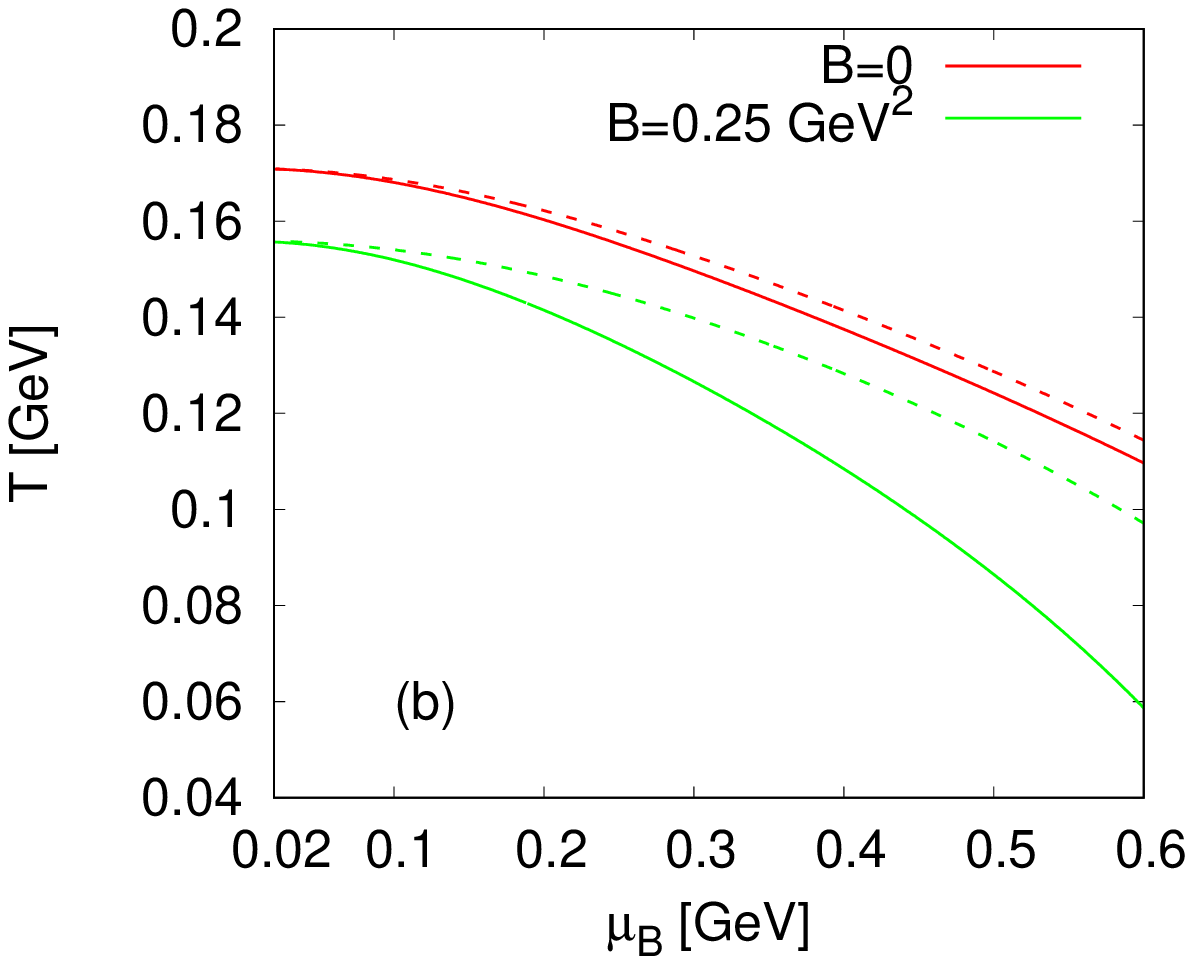}}
\subfloat{\includegraphics[scale=0.45]{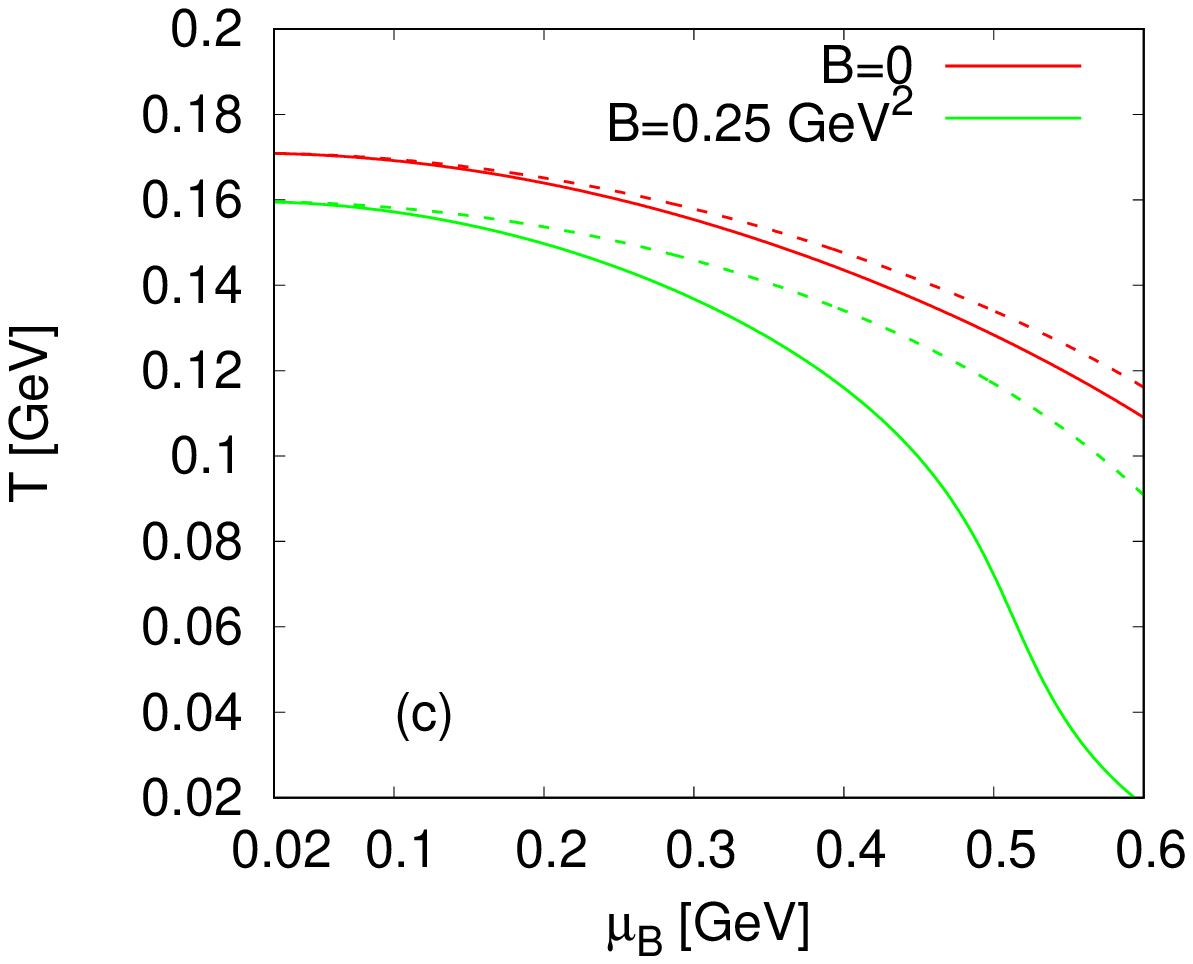}}
\caption{}{Chemical freeze-out curve determined by (a) $E/N\simeq$ 1 GeV,
(b) $n_{B}+n_{\bar B}\simeq 0.12$ $fm^{-3}$ and (c) $s/T^3\simeq 7$  with (dashed line) and
without (solid line) charge conservation.}
\label{Fig.1}
\end{figure}

\section {RESULTS AND DISCUSSION}

\subsection{ Quantities related to the conserved charges along the freeze-out curve}

\begin{figure*}[!hpt]
\begin{center}
\leavevmode
\epsfysize=6truecm \vbox{\epsfbox{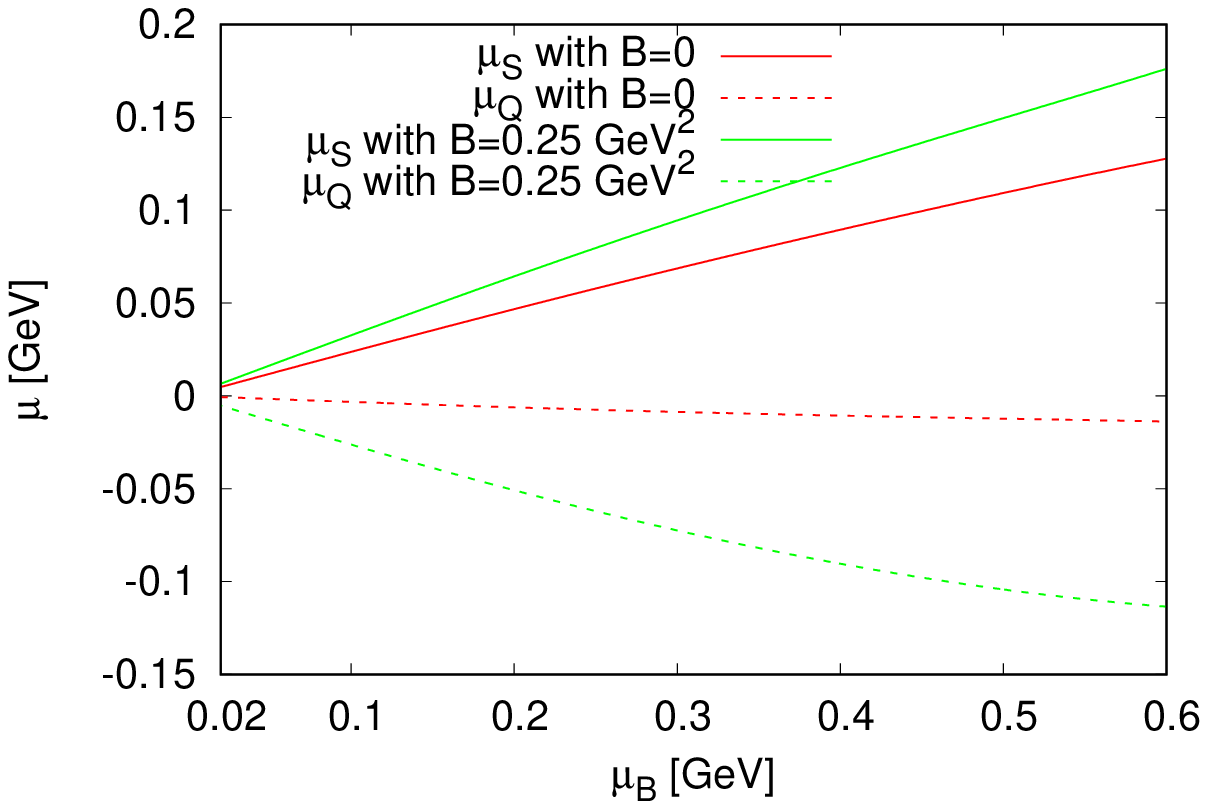}}
\end{center}
\caption{}{Strangeness and electric charge chemical potential along the freeze-out 
curve $E/N\simeq$ 1 GeV}
\label{Fig.1}
\end{figure*}  

Fig.2 represents $\mu_{S}$ and $\mu_{Q}$ as a function $\mu_{B}$ along the freeze-out 
curve described in Fig.1a. The solid line in Fig.2 shows the variation of $\mu_{S}$
and the dashed line shows the variation of $\mu_{Q}$ along the freeze-out curve. $\mu_{S}$
at nonzero $B = 0.25$ $GeV^2$ is always larger than $\mu_{S}$ without magnetic field. At 
higher $\mu_{B}$ there are more baryons ( protons and neutrons)
in the system. Imposing charge conservation, i.e $B/Q\simeq 2.52$, one needs negative 
$\mu_{Q}$. $\mu_{Q}$ is of the order 0.1 GeV at higher $\mu_{B}$ at $B = 0.25$ $GeV^2$
and it is of the order 0.01 GeV at zero magnetic field.

The variation of the normalized conserved net charge density with   
$\mu_{B}$ has been shown along the freeze-out curve (Fig.1a) in Fig.3. Fig.3a shows the variation
of $n_{Q}/T^3$ with $\mu_{B}$ along the freeze-out curve. The solid curve 
represents the normalized net electric charge density without charge conservation and 
the dash-dotted curve represents the normalized net electric charge density 
with charge conservation. 
At zero B and without charge conservation, for mesons $\mu_{B}=\mu_{S}=\mu_{Q}=0$.
So the contribution towards the net electric charge density is exactly canceled due
to the charged meson and its anti-particle. But $\mu_{B}$ is nonzero for baryons, so
the contribution towards the net electric charge density comes from the electrically charged 
baryons and anti-baryons. The most dominant positive contribution comes from the lowest mass 
charged baryon i.e proton and negative contribution comes from anti-proton. As $\mu_{B}$ 
increases, the number density of proton (baryons) increases and the number density of anti protons 
(anti-baryons) decreases, so the net electric charge density increases. At zero B and 
with charge conservation, the contribution coming from the electrically charged meson and 
its anti-particle don't cancel due to nonzero $\mu_{S}$ and/or $\mu_{Q}$. Hence 
there is extra contribution from mesons with 
charge conservation at $B = 0$. Also, the freeze-out temperature is higher with charge
conservation compared to without charge conservation at zero B. So, the number density
increases at higher temperature. So, the net electric charge density with charge conservation
increases slightly compared to without charge conservation at $B = 0$. At nonzero B and 
without charge conservation, the meson and its anti-particle densities are same and 
hence net electric charge density due to mesons and their anti-particles is zero. 
However, the number density of electrically charged baryons and anti baryons 
increases at nonzero magnetic field and the net electric charge density is large at 
nonzero B compared to zero B. But, with charge conservation at nonzero B,
the number density of protons decreases, hence the net electric charge density 
decreases. However, the net electric charge density with charge conservation
is two times larger at $B = 0.25$ $GeV^2$ compared to $B = 0$ at 
higher $\mu_{B}$.  

The chemical freeze-out temperature decreases due to the IMC effect at nonzero B.
However, if one assumes the freeze-out parameters 
at nonzero B are same as at $B = 0$, then one could use the fitted freeze-out parameters
of $T$, $\mu_{B}$, $\mu_{S}$ and $\mu_{Q}$ obtained in Eq.3 and Eq.4  
as given below \cite{redlich} at nonzero B. 

\begin{equation}
T = a - b{{\mu_{B}}^2} - c{{\mu_{B}}^4}
\end{equation}

where $a=(0.166\pm0.002)$ GeV, $b=(0.139\pm0.016)$ $GeV^{-1}$ and $c=0.053\pm0.021)$
$GeV^{-3}$

\begin{equation}
\mu_{X}(\sqrt{s_{NN}}) =\frac{d}{1+e{\sqrt{s_{NN}}}}
\end{equation}

Here X is the chemical potential for the different conserved charges and the 
corresponding values of d and e are given in the table below. 

\begin{table}[h!]
\centering
\caption{Parameterization of chemical potential $\mu_{X}$ along the freeze-out curve}
\label{tab:table1}
\begin{tabular}{l|c|r}
    X & d[GeV] & e[${GeV}^{-1}$]\\
    \hline
    B & 1.308(28) & 0.273(8)\\
    S & 0.214 & 0.161\\
    Q & 0.0211 & 0.106
  \end{tabular}
\end{table}

It is clear that the normalized electric charge density at $B = 0.25$ $GeV^2$ with 
charge conservation (dash-dotted line with the IMC effect) is far away from the 
charge density obtained using 
the fitted fixed chemical freeze-out (CF) parameters (dotted line). 
     
\begin{figure}
\centering
\subfloat{\includegraphics[scale=0.44]{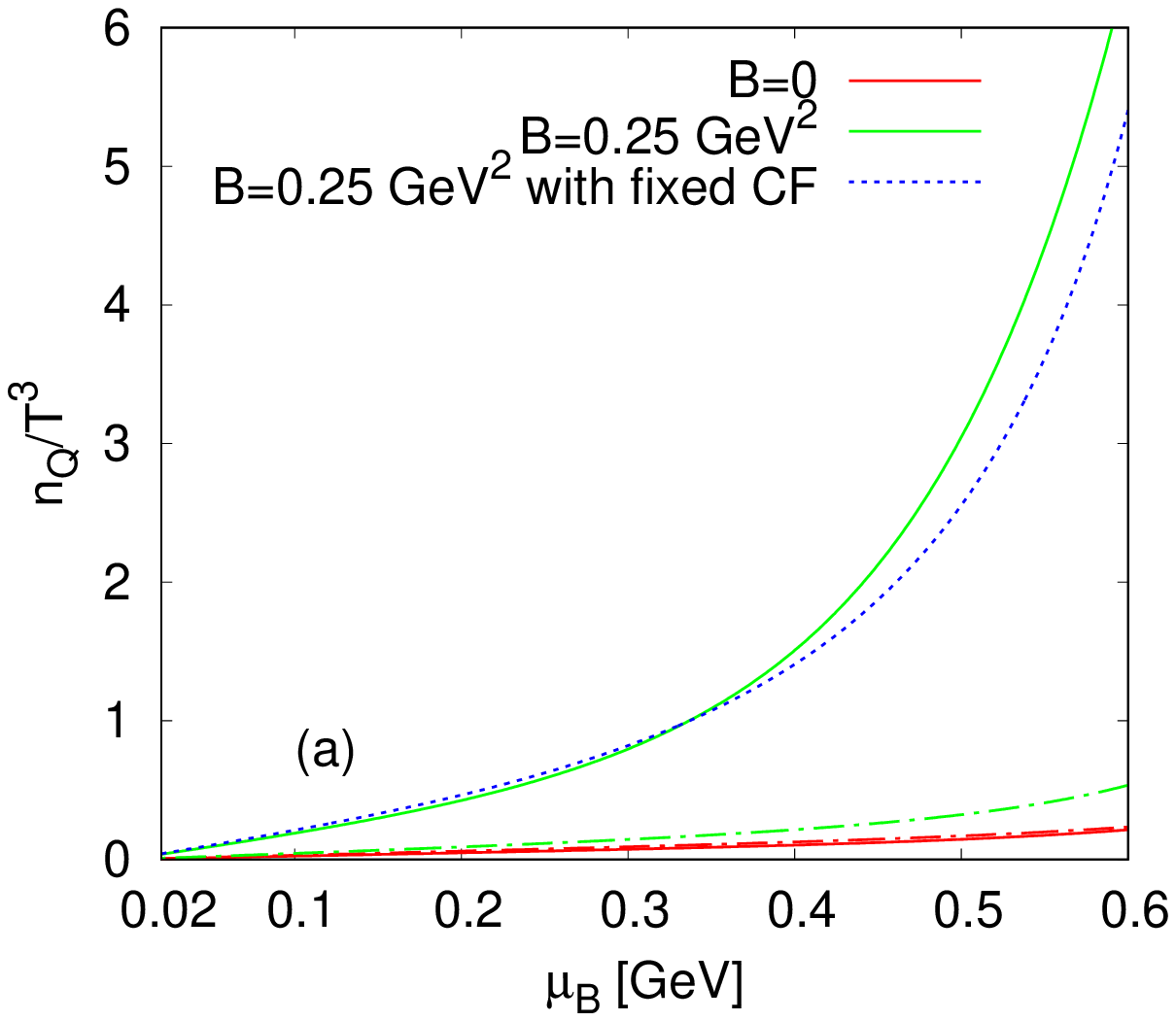}}
\subfloat{\includegraphics[scale=0.44]{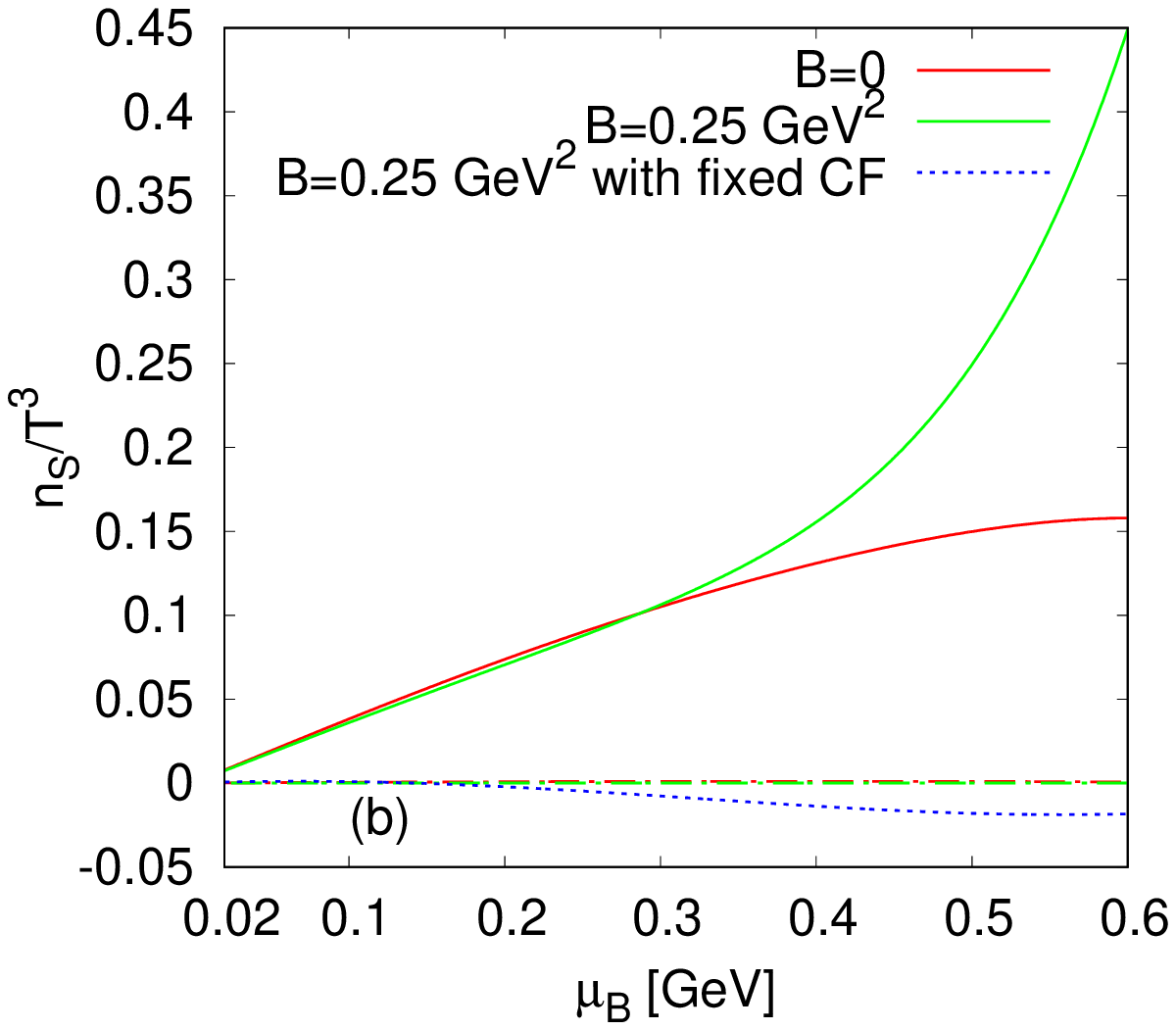}}
\subfloat{\includegraphics[scale=0.44]{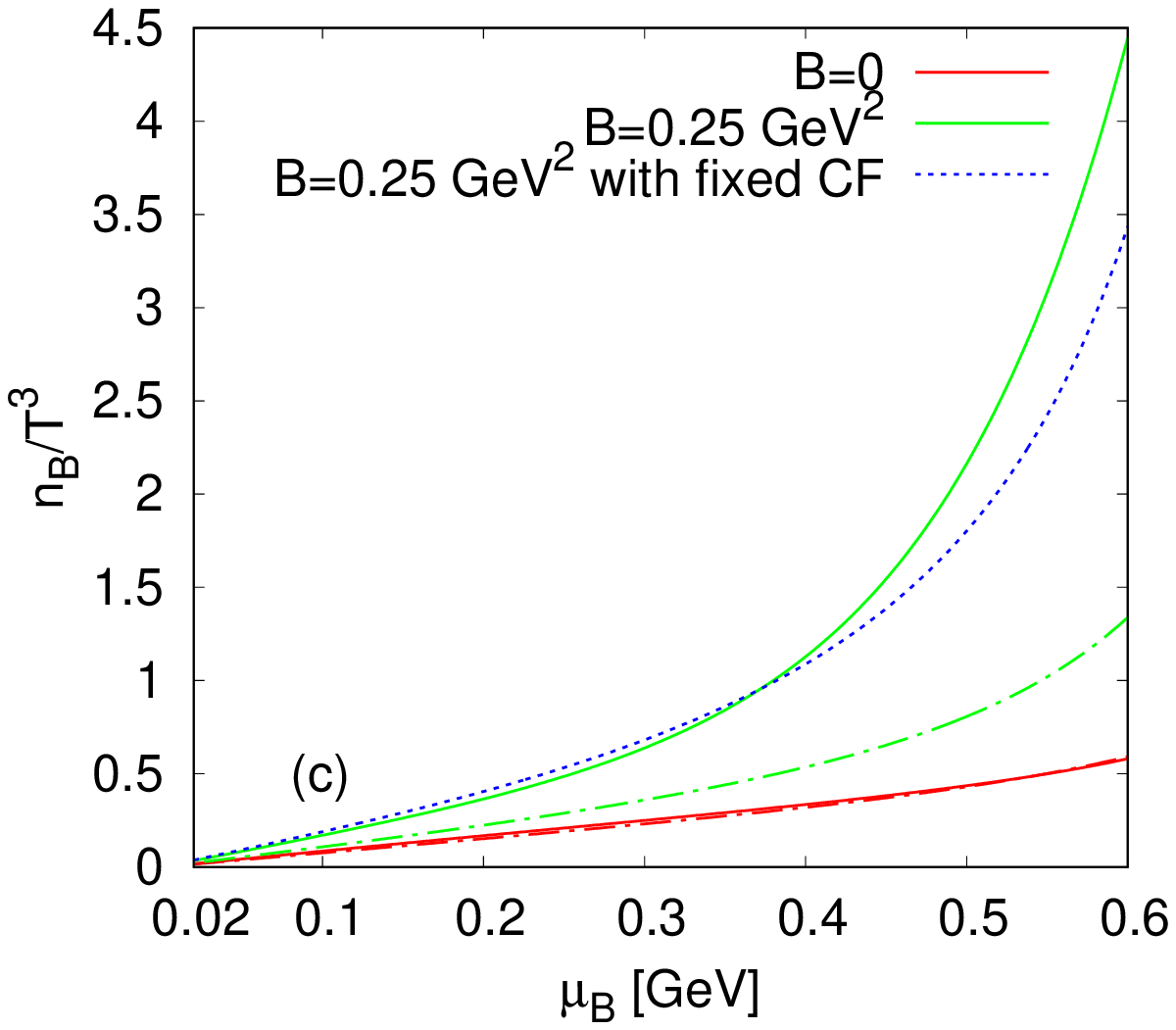}}
\caption{}{Normalized conserved net charge density along the freeze-out curve $E/N\simeq$ 1 GeV
at $B = 0$, $B = 0.25$ $GeV^2$ and $B = 0.25$ $GeV^2$ with fixed chemical 
freeze-out (CF) parameters obtained at zero B (dotted line). Here the solid line 
corresponds to the net charge density without charge conservation and the 
dash-dotted line corresponds to the net charge density with charge conservation.}
\label{Fig.3}
\end{figure}

Fig.3b shows the variation of ${n_{S}}/{T^3}$ with $\mu_{B}$ along 
the freeze-out curve. The solid curve shows ${n_{S}}/{T^3}$ without
charge conservation and the dash-dotted curve represents with charge conservation.
The normalized net strange charge density is zero when the charge conservation is taken 
into account, i.e net $S=0$ for zero and nonzero B. At $B = 0$ and without charge 
conservation, the contribution towards the net strange density is zero due to the strange 
mesons and their anti-particles. So the only strange 
baryons and their anti-particles contribute towards the net strange density. Here one shows 
the difference between anti-strange and strange particle contribution 
in the figure. The normalized net strange charge density is larger at nonzero B 
compared to zero B at higher $\mu_{B}$ without charge conservation. However, they 
are almost same at lower $\mu_{B}$. This is due to the fact that the dominant 
contribution towards the strange density comes from $\Lambda$ particle and 
their anti particles which is electrically neutral and magnetic field does
not affect neutral particle production. But at higher $\mu_{B}$, the 
production of electrically charged strange baryons like ${\Sigma}^{+}$ and 
${\Sigma}^{-}$ increases in the presence of the magnetic field. 
So, the normalized net strange charge density is larger at nonzero B compared 
to zero B at higher $\mu_{B}$. The net strangeness 
density for the fitted chemical freeze-out parameters ( dotted line) at $B = 0.25$ $GeV^2$ 
does not match with the strangeness density with charge 
conservation (dash-dotted line with the IMC effect).

Fig.3c shows the variation of ${n_{B}}/{T^3}$ with $\mu_{B}$ along
the freeze-out curve. One can see at $B = 0$, ${n_{B}}/{T^3}$ is almost the same with and without 
charge conservation. However, at nonzero B, this is not the same with and without 
charge conservation. ${n_{B}}/{T^3}$ is always larger at nonzero B compared to 
zero B, because more electrically charged baryons ($\Delta$ particles and protons) are 
produced in nonzero magnetic field. The production of ${\Delta}^{++}$ particles increases 
compared to proton in nonzero B due to higher spin 3/2 of ${\Delta}^{++}$ and also higher 
electric charge.
The net baryon density with charge conservation decreases compared to 
no charge conservation in nonzero B. However, the conservation law
demands $B/Q\simeq 2.52$ or in other words ${n_B/{n_Q}}\simeq 2.52$. As one can see,
from Fig.3a and Fig.3c, this relation always holds good at any $\mu_B$ with charge 
conservation. I have also 
compared ${n_{B}}/{T^3}$ from 
the fitted chemical freeze-out parameters at $B = 0.25$ $GeV^2$ (dotted line), but 
this is away from the curve at $B = 0.25$ $GeV^2$ with charge conservation (dash-dotted 
line with the IMC effect).      

\subsection{Fluctuations and correlations along the freeze-out curve}

The fluctuations and the correlations are given by the diagonal and off diagonal 
components of susceptibility. These are defined by

\begin{equation} 
\chi_{xy}^{ij}=\frac{\partial^{i+j}\left(\sum_{k}{P_{k}/T^4}\right)}
{{\partial({\frac{\mu_{x}}{T}})^i}{\partial({\frac{\mu_{y}}{T}})^j}}
\end{equation}

These are important characteristics of any physical system. They
provide essential information about the effective degrees of freedom 
and their possible quasi-particle nature. They determine the response of the 
system to small external forces. The fluctuations are related to the phase transitions
and the fluctuations of all length scales are relevant at a critical point. The study of 
the fluctuations in heavy-ion collision should lead to a rich set of 
phenomena and is an essential tool for the experimental exploration
of the QCD critical point and the first order phase transition. In relativistic 
heavy ion collisions, the bulk fluctuations are observed by the 
event-by-event analyses. Here, a given observable
is measured on an event-by-event basis and the fluctuations are 
studied over the ensemble of the events. The fluctuations should diverge 
around a critical point (here finite due to finite size of the system created
in heavy-ion collisions). The fluctuations of the conserved charges in heavy-ion 
collisions was discussed in the context of the net electric charge fluctuations 
\cite {asakawa,koch}. The fluctuations in the net electric charge are
suppressed in the QGP phase compared to the hadronic gas phase since the quarks
carry fractional electric charges. It also has been pointed out that
the correlation between baryon number and strangeness is stronger in the QGP 
phase compared to the hadronic gas phase since the strange quarks have nonzero
baryon number in the QGP phase \cite{koch2}. But, in the hadronic gas phase, this 
correlation decreases since strange mesons don't carry the baryonic charge.

The electric charge susceptibility (2nd order) along the freeze-out curve is already 
shown in \cite{fuku2}. It has been shown that the electric charge susceptibility 
increases in nonzero B compared to zero B. Here I have presented all the fluctuations 
and the correlations along the freeze-out curve $E/N\simeq$ 1 GeV mentioned in Fig.1a.
One presents the strange susceptibility
and the baryon susceptibility (2nd order) along the freeze-out curve at $B = 0$ and 
$B = 0.25$ $GeV^2$ in Fig.4a and Fig.4b respectively. I also have compared the results from 
fitted chemical freeze-out parameters at $B = 0.25$ $GeV^2$. 

\begin{figure}
\centering
\subfloat{\includegraphics[scale=0.5]{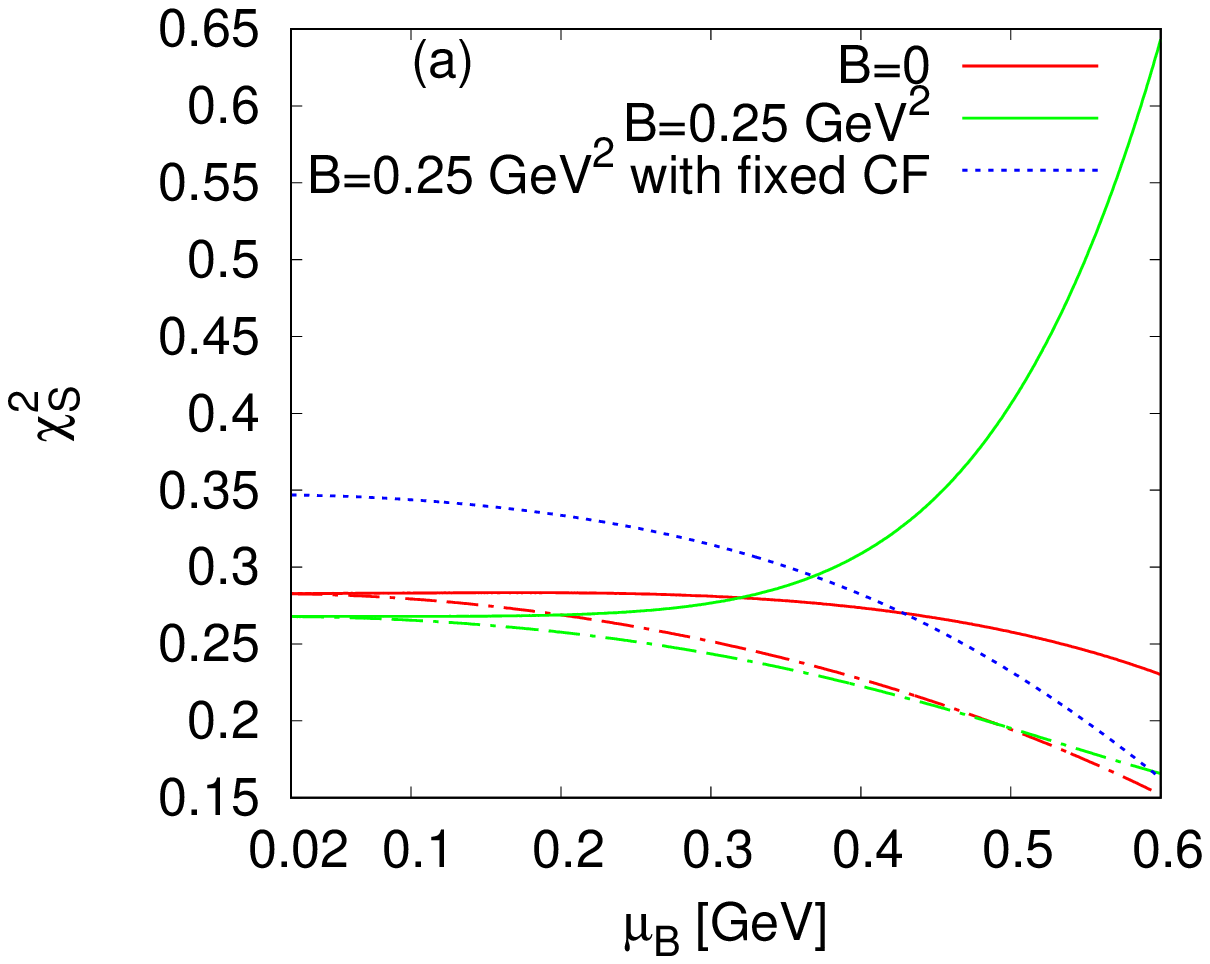}}
\subfloat{\includegraphics[scale=0.5]{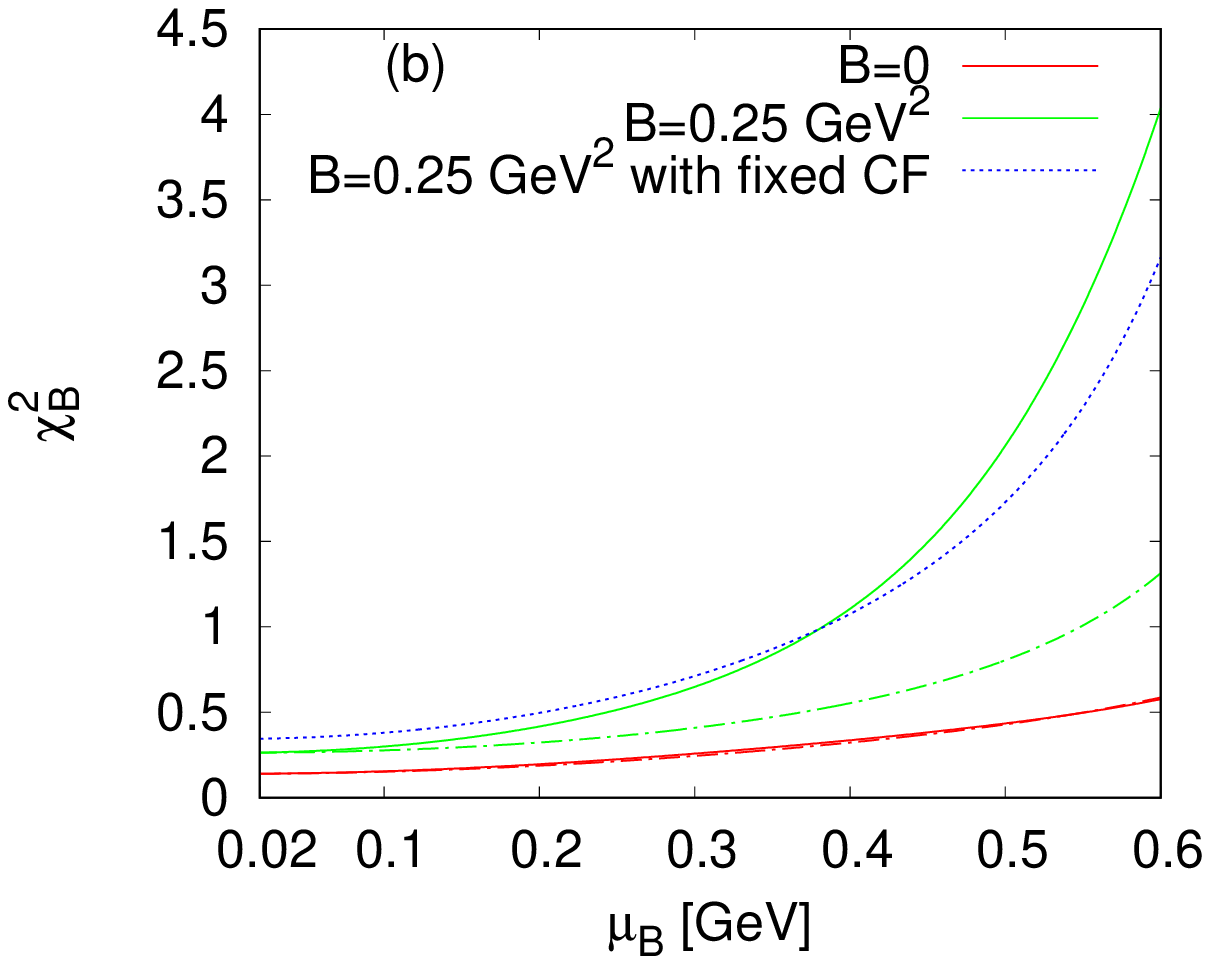}}
\caption{}{ (a) $\chi_{S}^2$ and (b) $\chi_{B}^2$ along the freeze-out curve at $B = 0$, 
$B = 0.25$ $GeV^2$ and $B = 0.25$ $GeV^2$ with fixed chemical freeze-out (CF) parameters
obtained at zero B (dotted line). Here the solid line
corresponds to the charge susceptibility without charge conservation and the
dash-dotted line corresponds to the charge susceptibility with charge conservation.}
\label{Fig.4}
\end{figure}

The most dominant contribution towards the strange susceptibility comes from the lowest 
mass strange particle i.e kaons.
At $B = 0$, $\chi_{S}^2$ is large without charge conservation (solid line) 
compared with charge conservation (dash-dotted line) along the freeze-out curve (Fig.4a). This 
is also true for nonzero B. So, the charge conservation diminishes the strangeness 
fluctuations along the freeze-out curve for zero and nonzero B. One can 
see $\chi_{S}^2$ with charge conservation at nonzero B is slightly below the curve at 
zero B because the effective mass of kaon increases at nonzero B. 
$\chi_{S}^2$ at $B = 0.25$ $GeV^2$ from the fitted parameters (dotted line) is very 
different than the results at nonzero B with charge conservation. 

At $B = 0$, $\chi_{B}^2$ is almost the same with (dash-dotted line) and without charge
conservation ( solid line) as shown in Fig.4b. At $B = 0.25$ $GeV^2$, the second order 
baryon susceptibility (the baryon fluctuations) is very large when there is no 
charge conservation (due to more protons production). However, when the charge 
conservation is taken into account, the baryon fluctuations decrease at 
nonzero B.  $\chi_{B}^2$ at $B = 0.25$ $GeV^2$ from 
the fitted parameters (dotted line) is very different than the results at 
nonzero B with the charge conservation. From above discussions, it is clear that if 
the freeze-out parameters at nonzero B are same as at zero B (i.e using zero B fitted parameters 
at nonzero B), the fluctuations in the conserved charges are different compared to the 
fluctuations at nonzero B with the charge conservation. So it is always important
to fix the strange and the electric charge chemical potentials 
using the conservation laws at nonzero B, rather than using zero B fitted parameters 
at nonzero B. This is also true for all the higher order fluctuations. From here 
onwards, I will not compare the results from the fitted parameters.     
   
\begin{figure}
\centering
\subfloat{\includegraphics[scale=0.45]{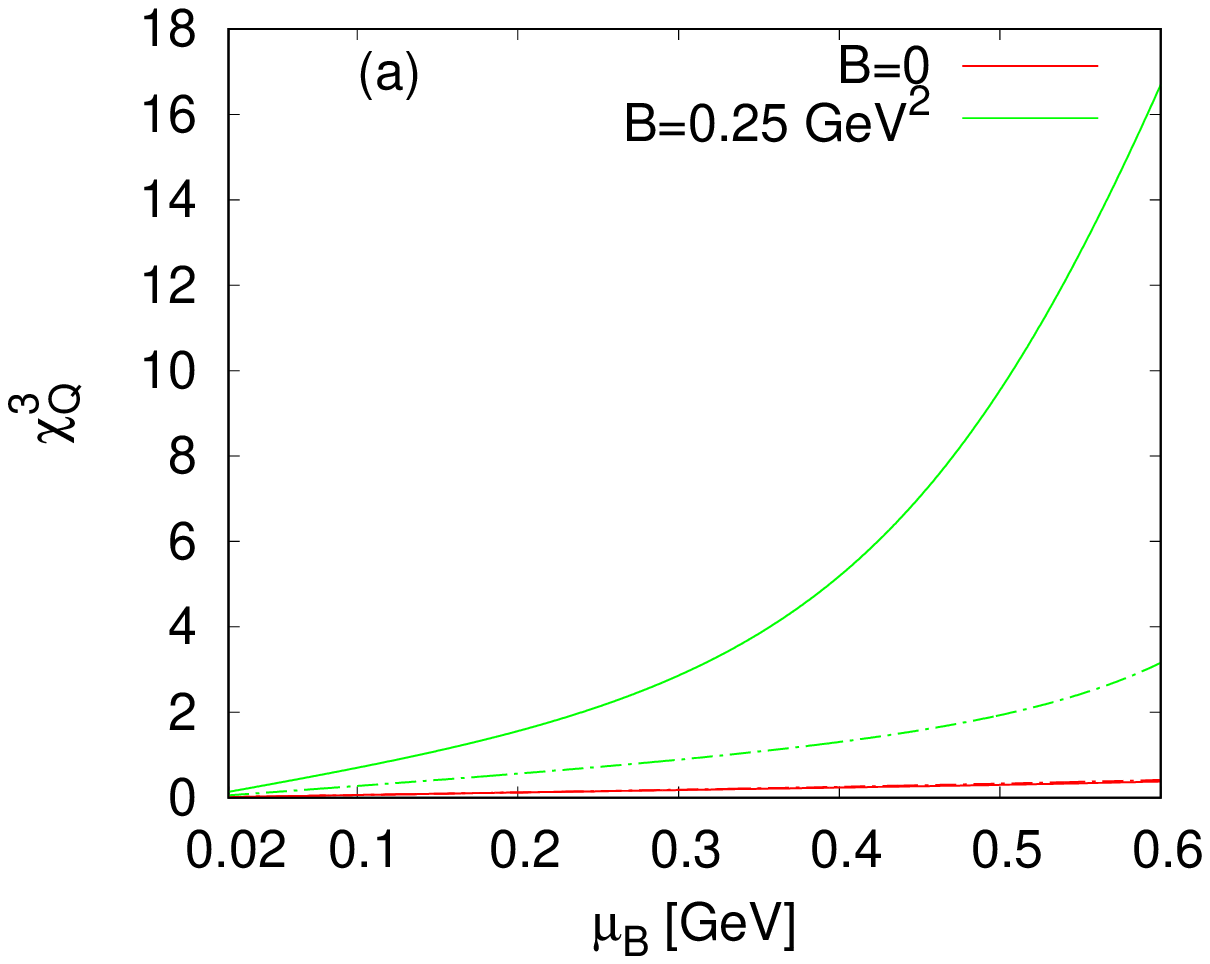}}
\subfloat{\includegraphics[scale=0.45]{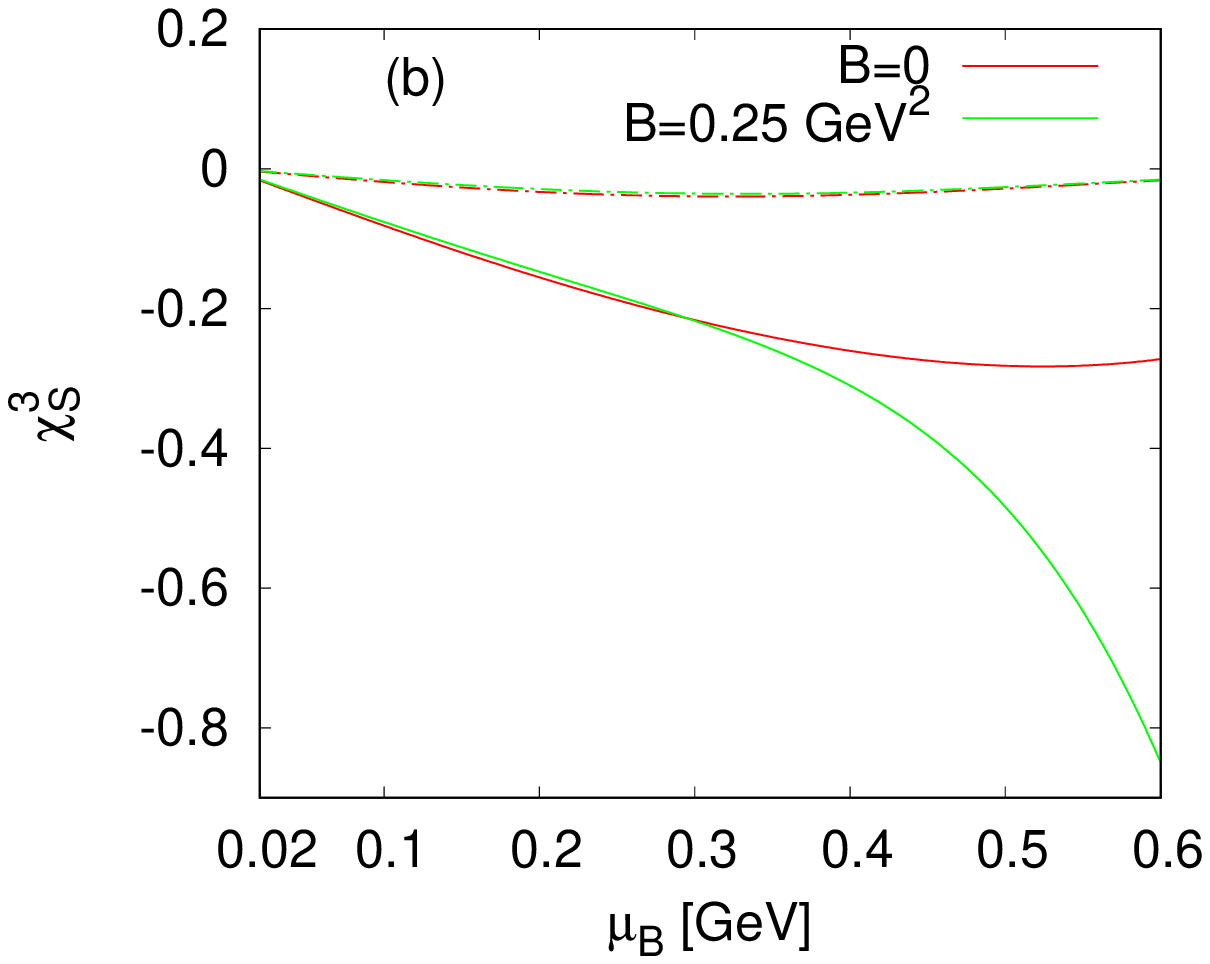}}
\subfloat{\includegraphics[scale=0.45]{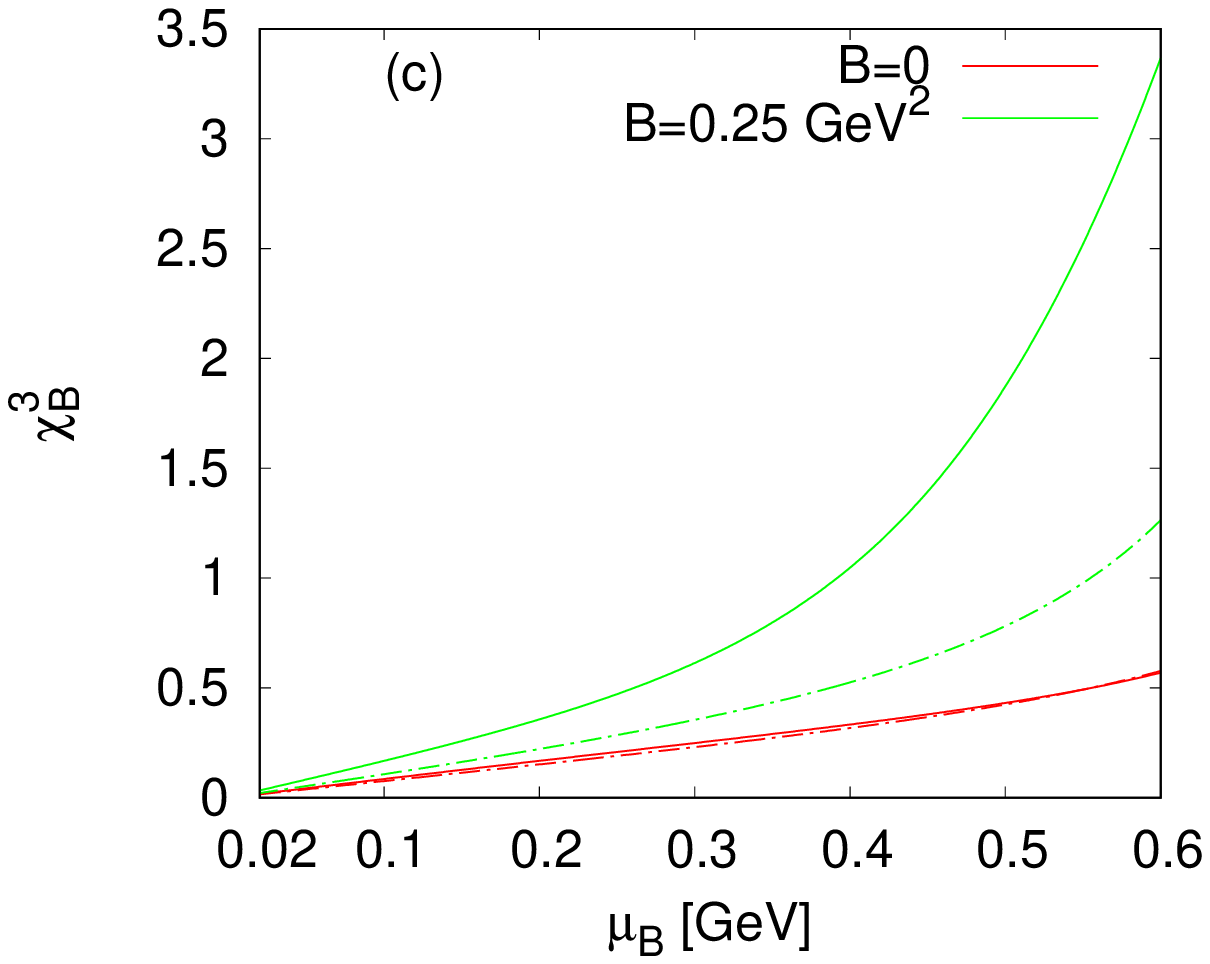}}
\caption{}{ (a) $\chi_{Q}^3$, (b) $\chi_{S}^3$ and (c) $\chi_{B}^3$ along the 
freeze-out curve with and without B. Here the solid line
corresponds to the charge susceptibility without charge conservation and the
dash-dotted line corresponds to the charge susceptibility with charge conservation.}
\label{Fig.5}
\end{figure}

Fig.5 presents the variation of the third order susceptibilities corresponding to 
the conserved charges as a function of $\mu_{B}$ along the freeze-out curve. One can 
see the variation of the second and the third order susceptibilities are similar for 
the electric charge and the baryonic charge. However, the susceptibility (3rd order)
corresponding to the strange charge change in sign. $\chi_{S}^3$ (modulus value)
at zero and nonzero B decreases when the charge conservation is taken into account. 

Fig.6 shows the  variation of the fourth order susceptibilities corresponding to 
the conserved charges as a function of $\mu_{B}$ along the freeze-out curve.
The variation of even order susceptibilities (2nd and 4th order) corresponding 
to the conserved charges are similar. However, the values of the 4th order susceptibility
are generally larger compared to the 2nd order susceptibilities due to the larger weight 
for higher order.
Since all baryons have baryon number 1, $\chi_{B}^2\simeq\chi_{B}^4$ for zero B.
At nonzero B and with charge conservation, this relation approximately holds good.
However, $\chi_{B}^4$ is less than $\chi_{B}^2$ at nonzero B and without charge conservation.

\begin{figure}
\centering
\subfloat{\includegraphics[scale=0.45]{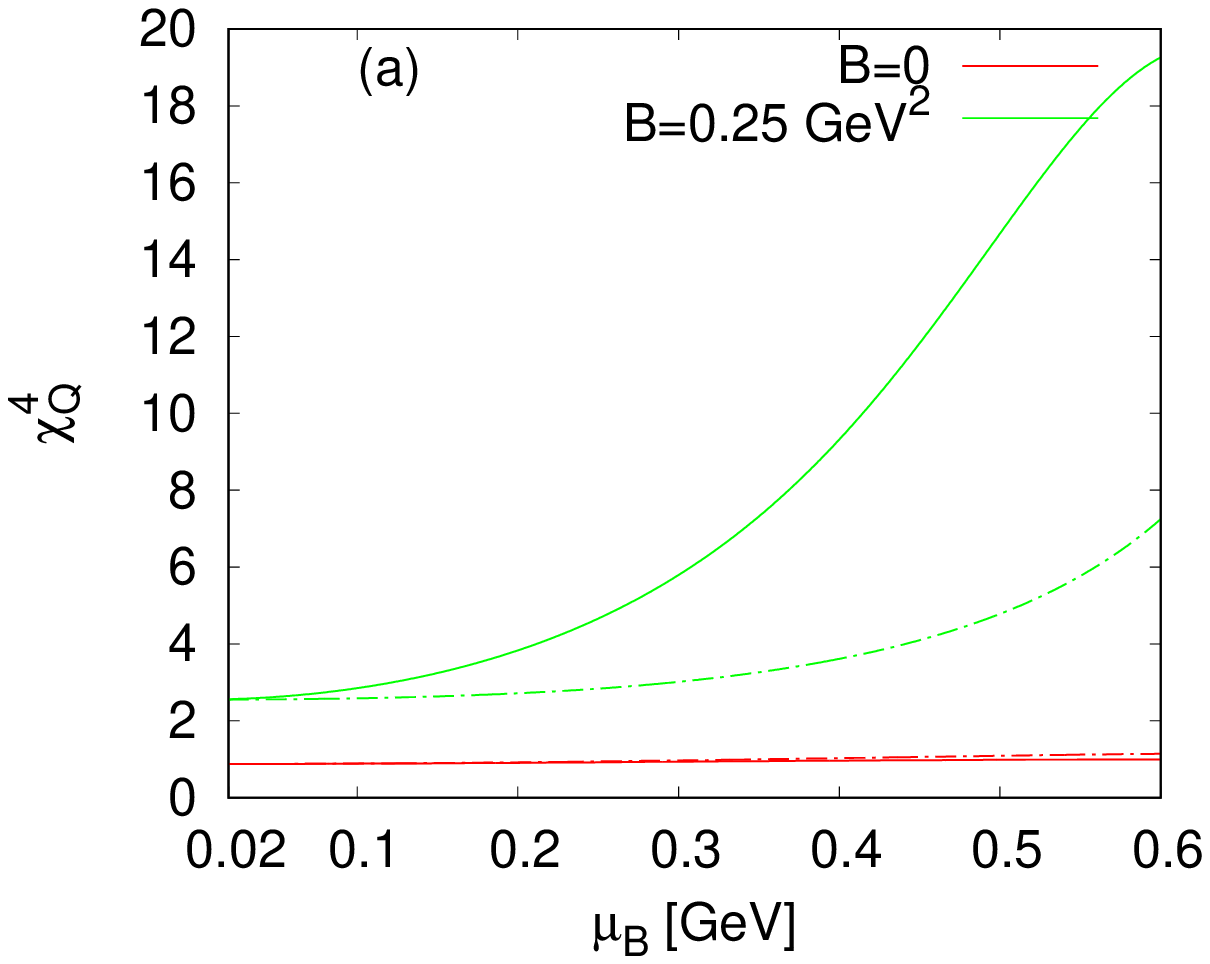}}
\subfloat{\includegraphics[scale=0.45]{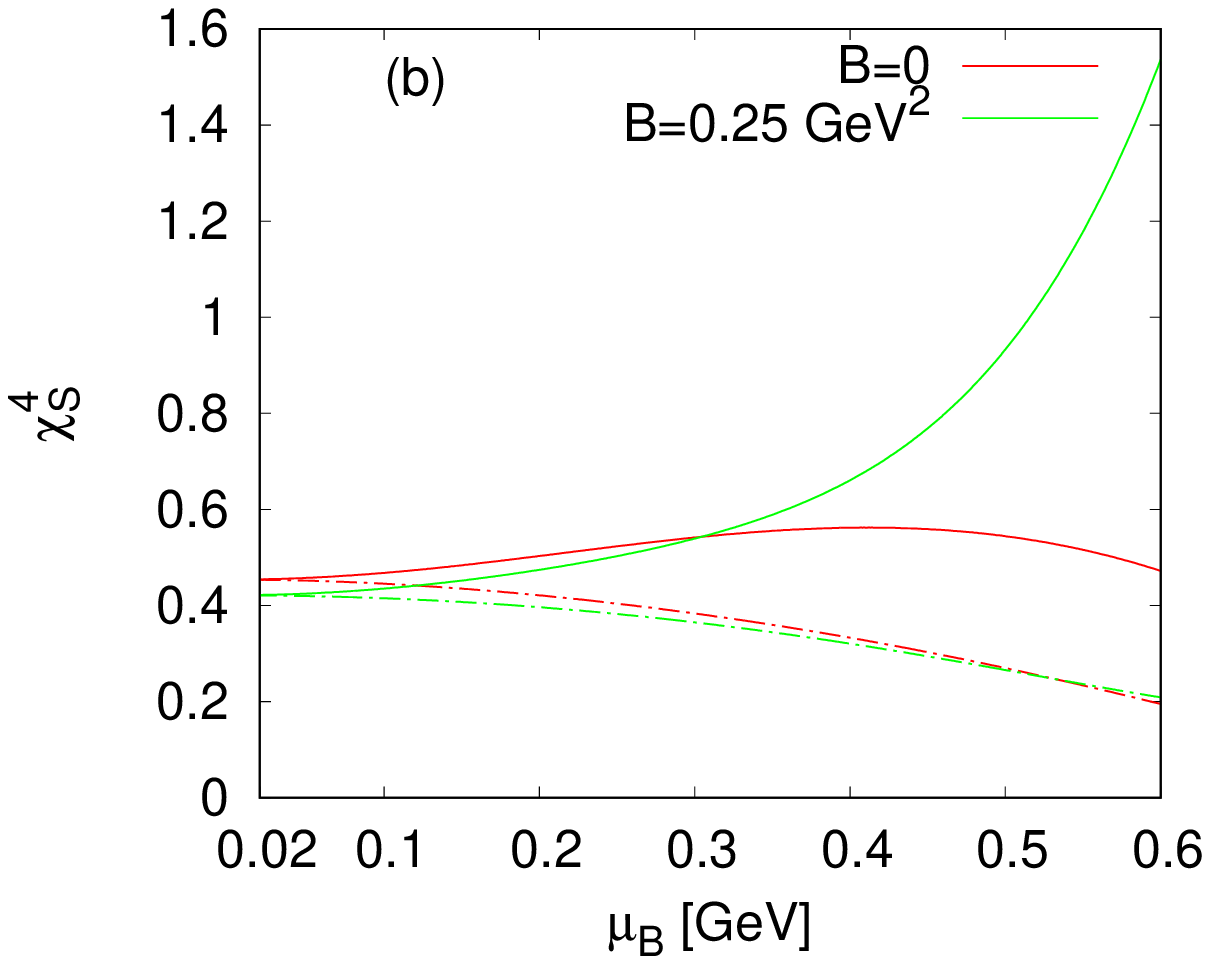}}
\subfloat{\includegraphics[scale=0.45]{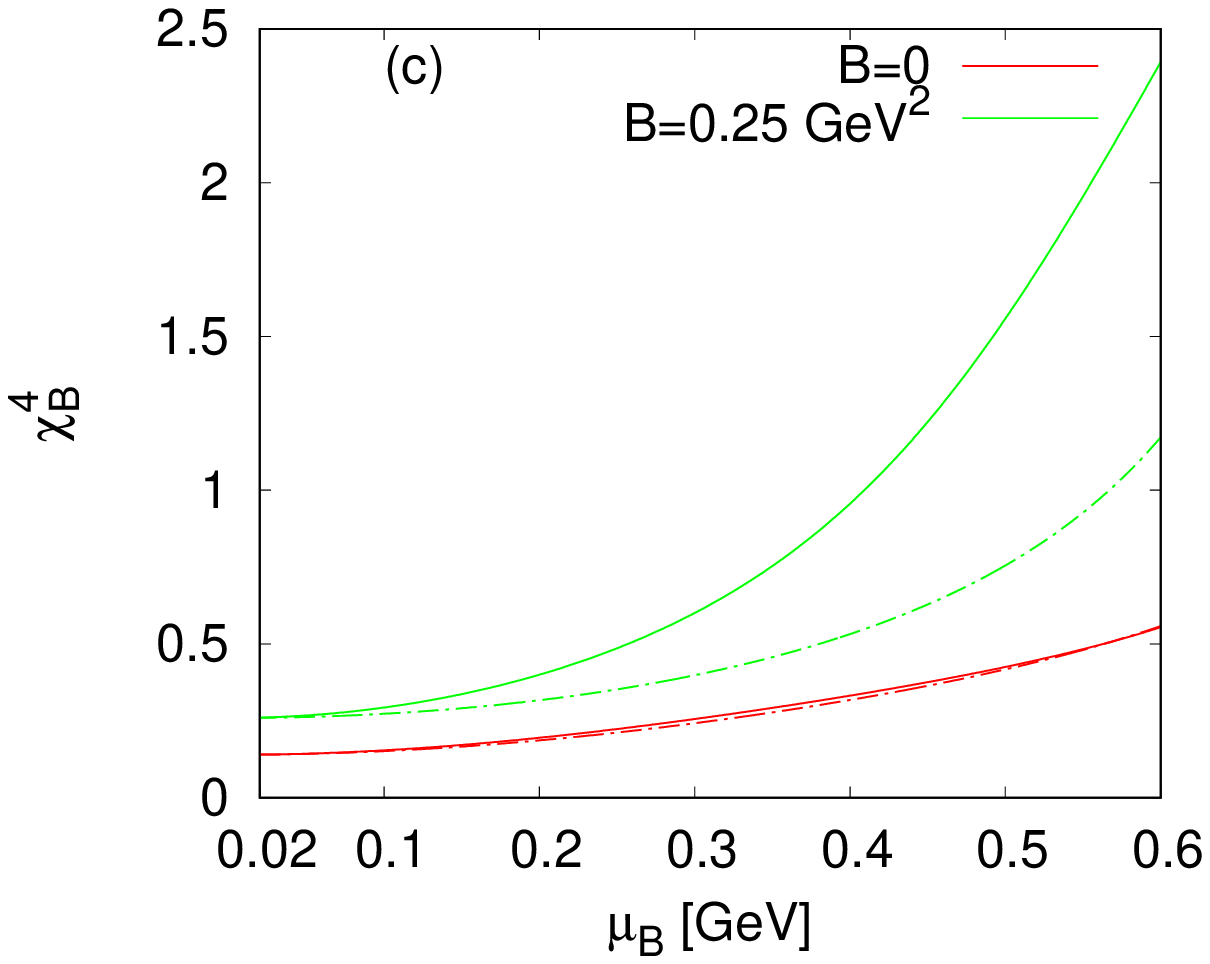}}
\caption{}{ (a) $\chi_{Q}^4$, (b) $\chi_{S}^4$ and (c) $\chi_{B}^4$ along the
freeze-out curve with and without B. Here the solid line
corresponds to the charge susceptibility without charge conservation and the
dash-dotted line corresponds to the charge susceptibility with charge conservation.}
\label{Fig.6}
\end{figure}

The conserved charge correlations have been shown along the freeze-out curve in Fig.7. 
Fig.7a presents $\chi_{QS}$ as a function of $\mu_{B}$ along the freeze-out curve.
The dominant contribution towards the strangeness and the electric charge correlation 
comes from kaons (same nature of corresponding charge). At zero B, the correlation 
decreases towards higher $\mu_{B}$, because higher $\mu_{B}$ corresponds to lower 
freeze-out temperature and the number of kaons decreases due to decrease in temperature.
 The most important thing here to note that the correlation increases with charge 
conservation compared to no charge conservation at zero B. This is due to the fact 
that kaon density increases with charge conservation due to the increase
in chemical freeze-out temperature compared to no charge conservation. The correlation 
increases at nonzero B compared to zero B due to more production of the electric charges 
with nonzero strangeness (kaons) in nonzero magnetic field. The correlation also increases 
with charge conservation at nonzero B due to the increase in chemical freeze-out 
temperature. At nonzero B and higher $\mu_{B}=500$ $MeV$, one can see the correlation increases 
without charge conservation due to the increase of chemical freeze-out temperature as shown in
Fig.1a.   

\begin{figure}[h!]
\centering
\subfloat{\includegraphics[scale=0.45]{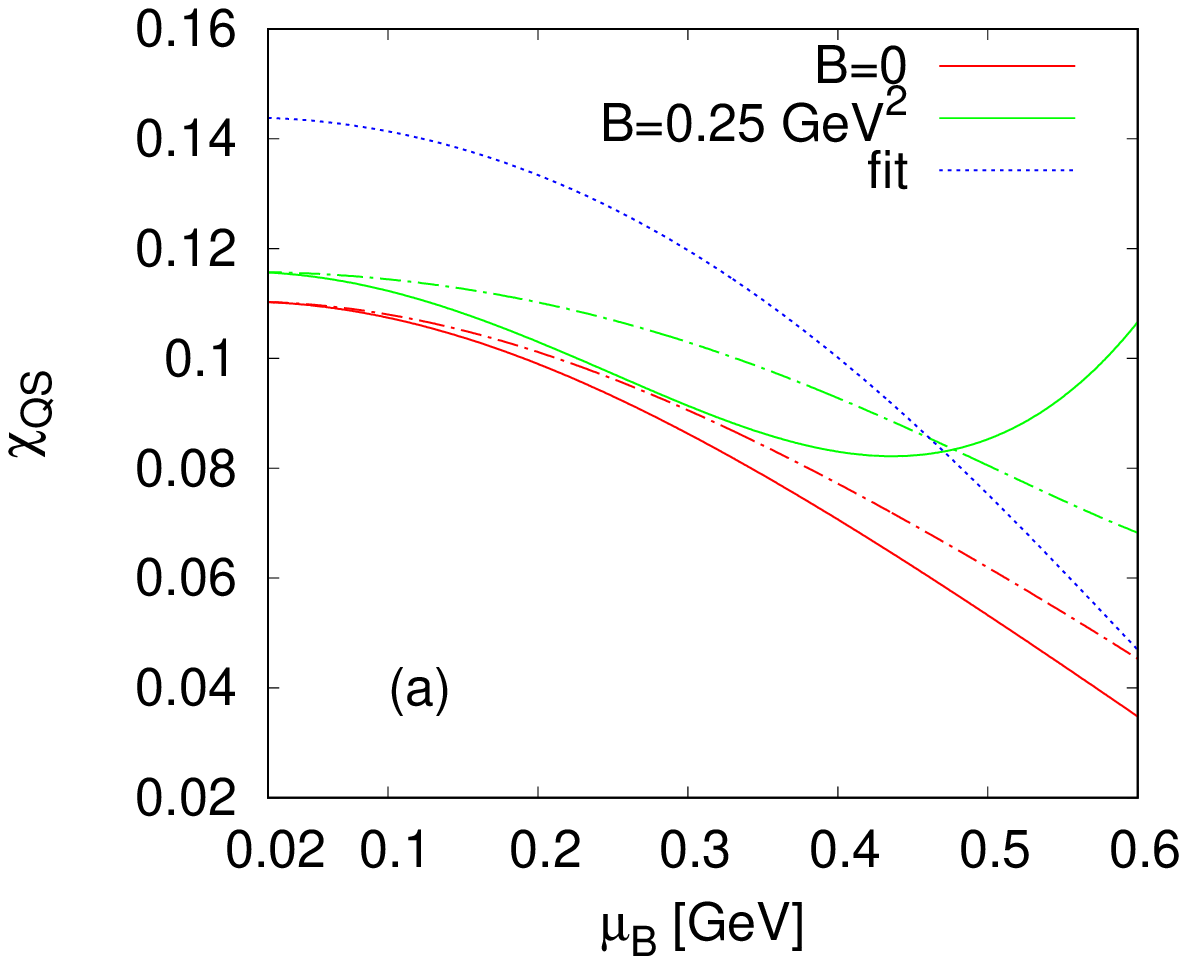}}
\subfloat{\includegraphics[scale=0.45]{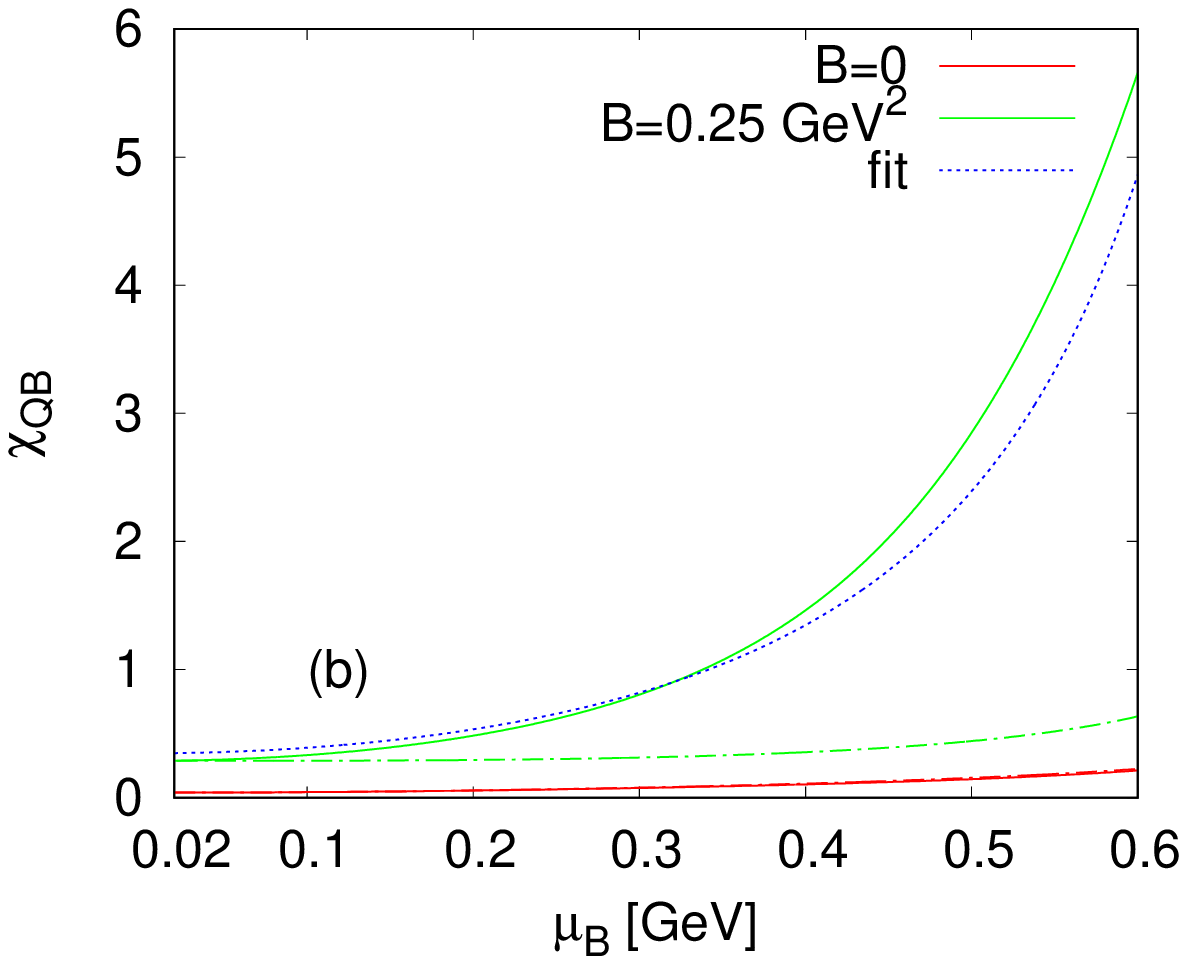}}
\subfloat{\includegraphics[scale=0.45]{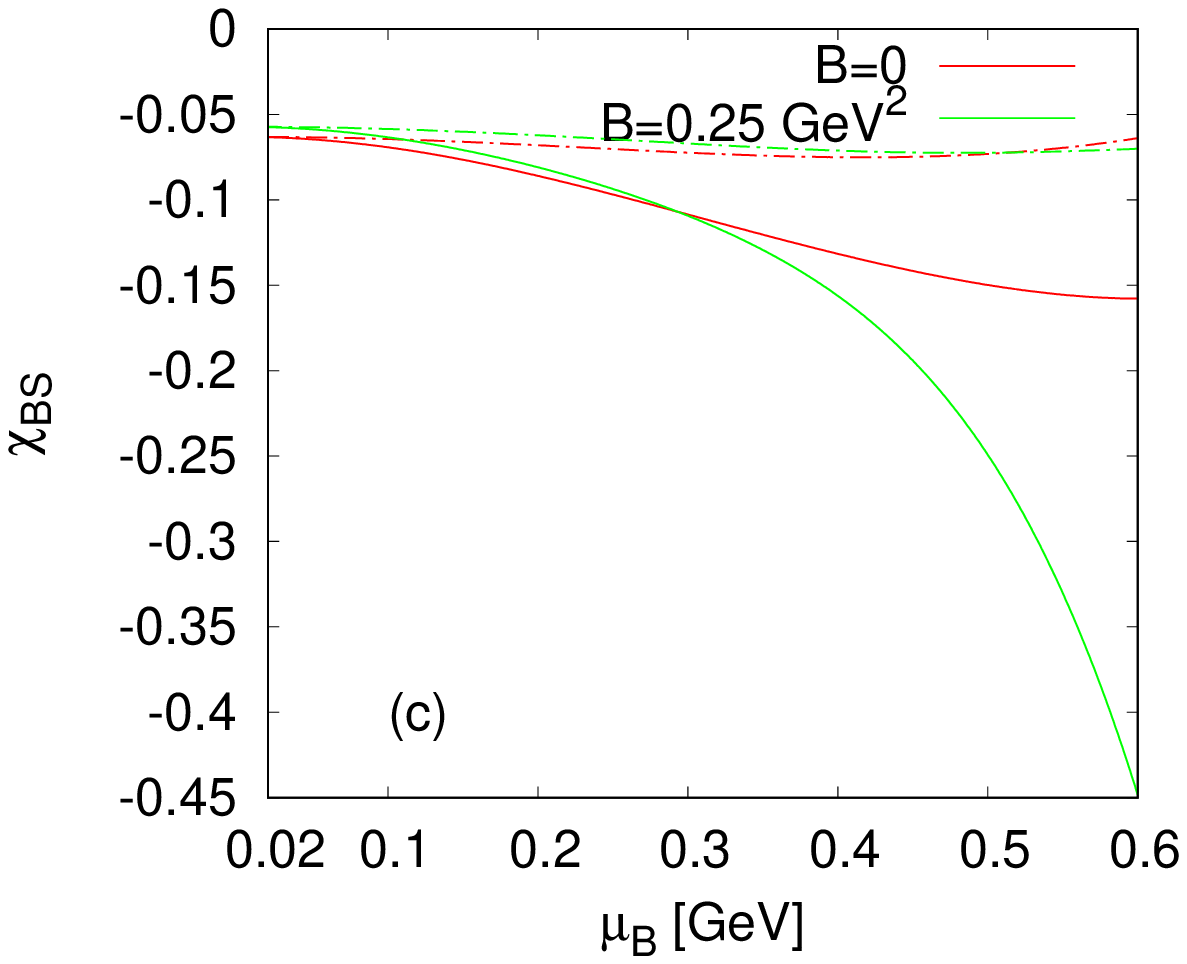}}
\caption{}{ (a) $\chi_{QS}$, (b) $\chi_{QB}$ and (c) $\chi_{BS}$ along the
freeze-out curve with and without B. Here the solid line
corresponds to the conserved charge correlations without charge conservation and the
dash-dotted line corresponds to the conserved charge correlations with charge conservation.}
\label{Fig.7}
\end{figure}

I have shown the variation $\chi_{QB}$ as a function of $\mu_{B}$ along the 
freeze-out curve in Fig.7b. At zero B, the electric and the baryonic charge correlation
is same with and without charge conservation. The correlation increases 
at nonzero B due to more production of protons and $\Delta^{++}$ without charge 
conservation. This correlation decreases when the charge conservation is 
taken into account.

Fig.7c presents the variation of $\chi_{BS}$ along the freeze-out curve.
The correlation is always negative because of the opposite nature of the corresponding 
charges. The dominant contribution comes from $\Sigma$ baryons. 
The correlation (modulus value) is larger without charge conservation than with 
charge conservation at zero and nonzero B. This is mainly due to more baryons 
production with nonzero strangeness when there is no charge conservation. 

\begin{figure}[h!]
\centering
\subfloat{\includegraphics[scale=0.45]{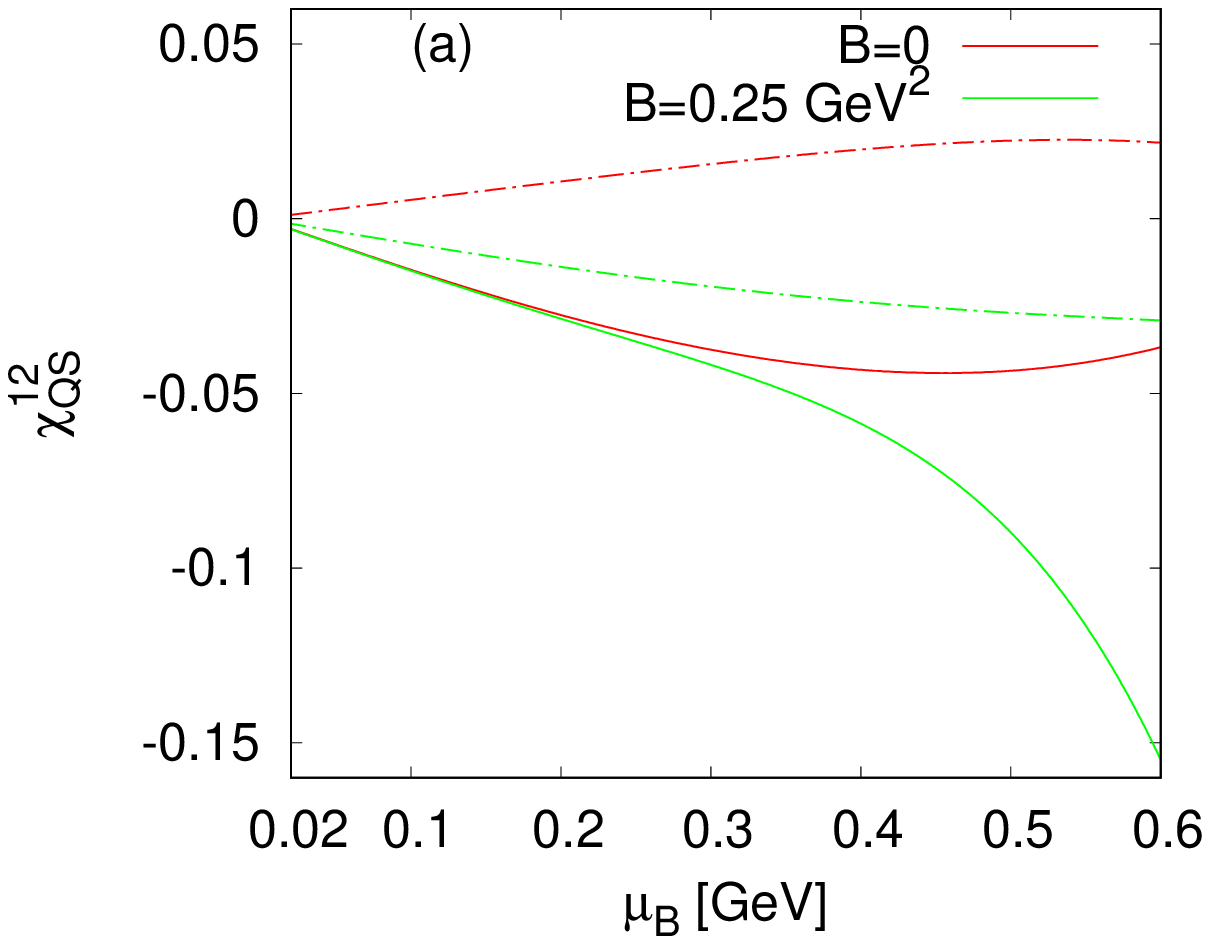}}
\subfloat{\includegraphics[scale=0.45]{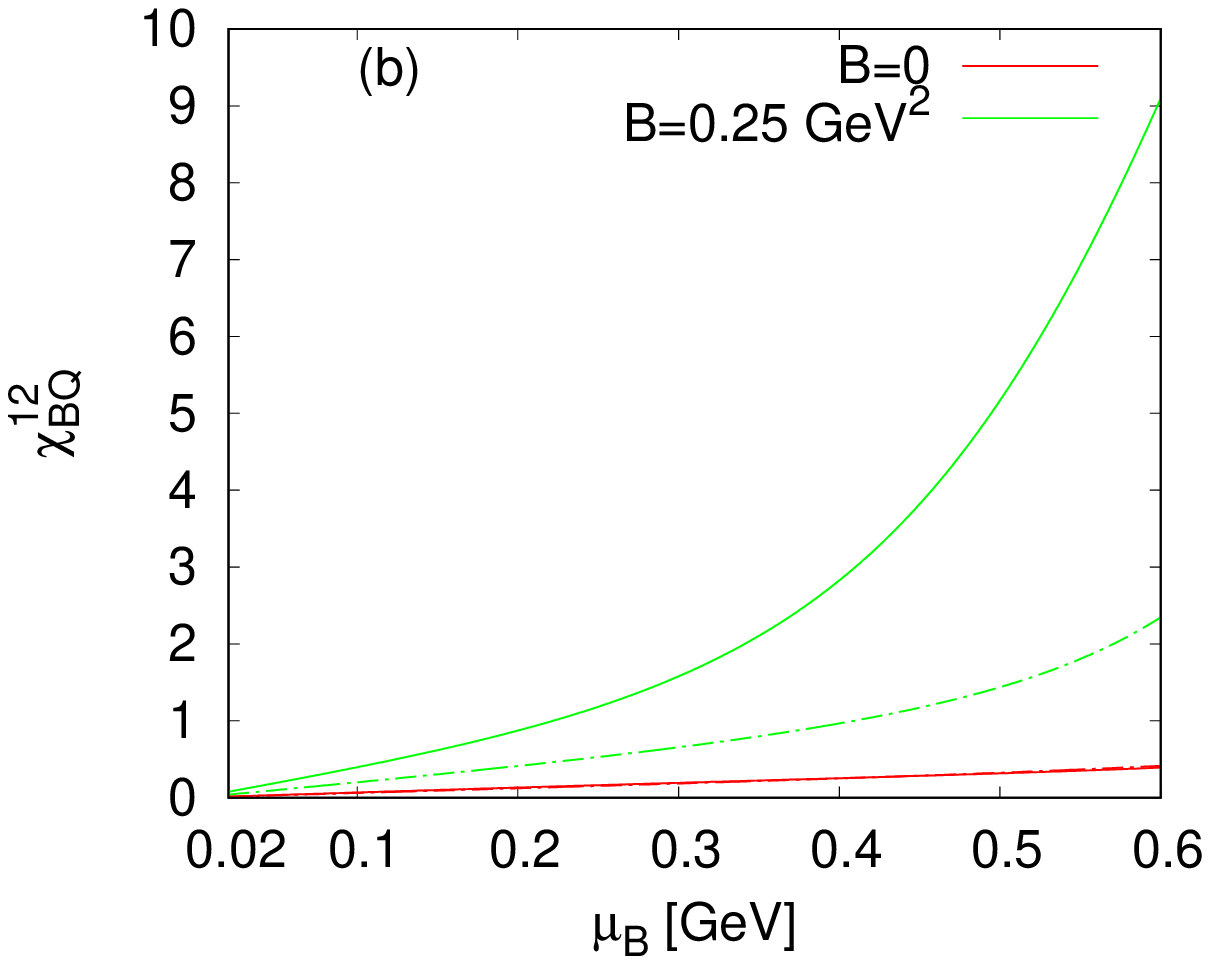}}
\subfloat{\includegraphics[scale=0.45]{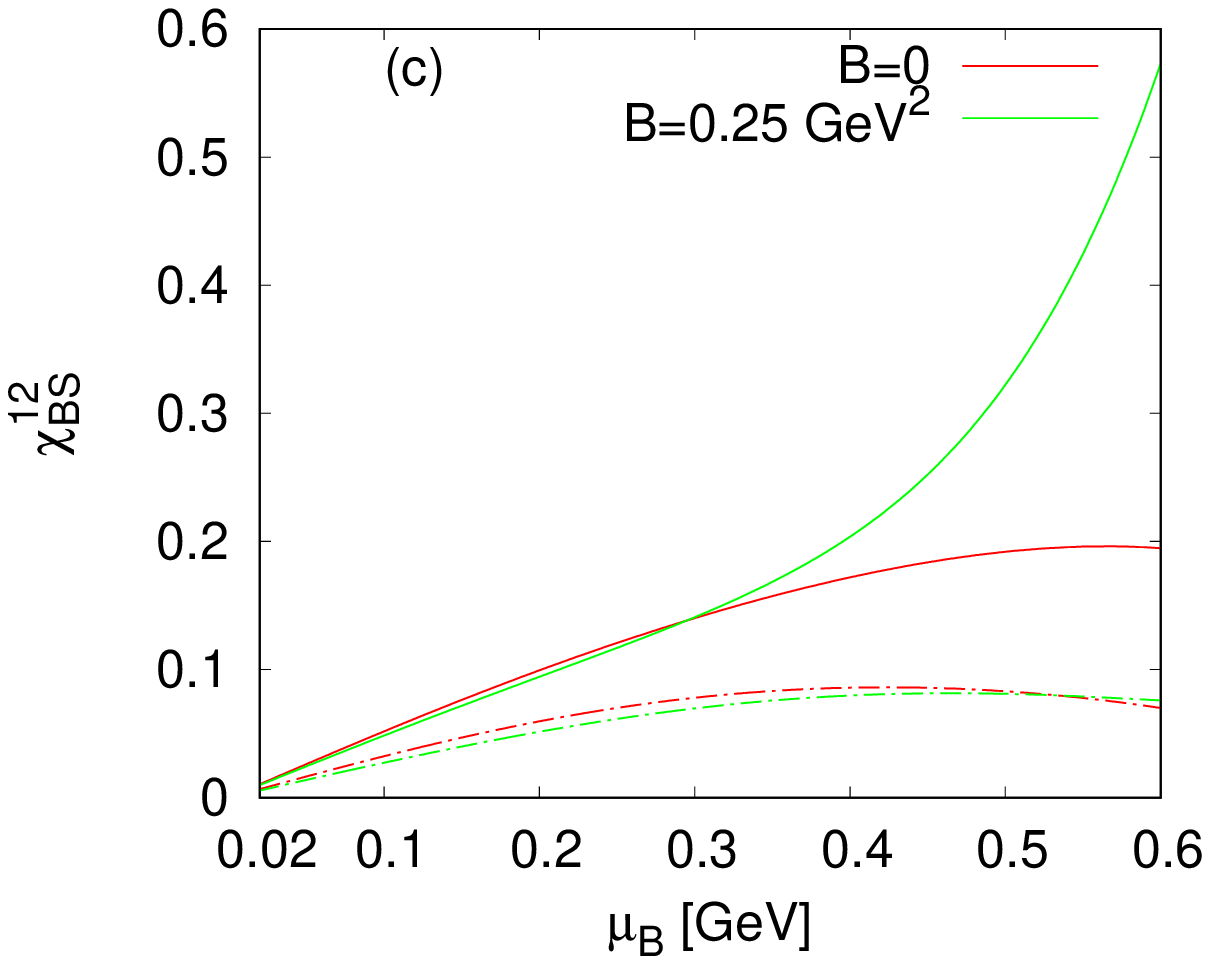}}
\caption{}{ (a) $\chi^{12}_{QS}$, (b) $\chi^{12}_{BQ}$ and (c) $\chi^{12}_{BS}$ along the
freeze-out curve with and without B. Here the solid line
corresponds to the conserved charge correlations without charge conservation and the
dash-dotted line corresponds to the conserved charge correlations with charge conservation.}
\label{Fig.8}
\end{figure}

\begin{figure}[h!]
\centering
\subfloat{\includegraphics[scale=0.45]{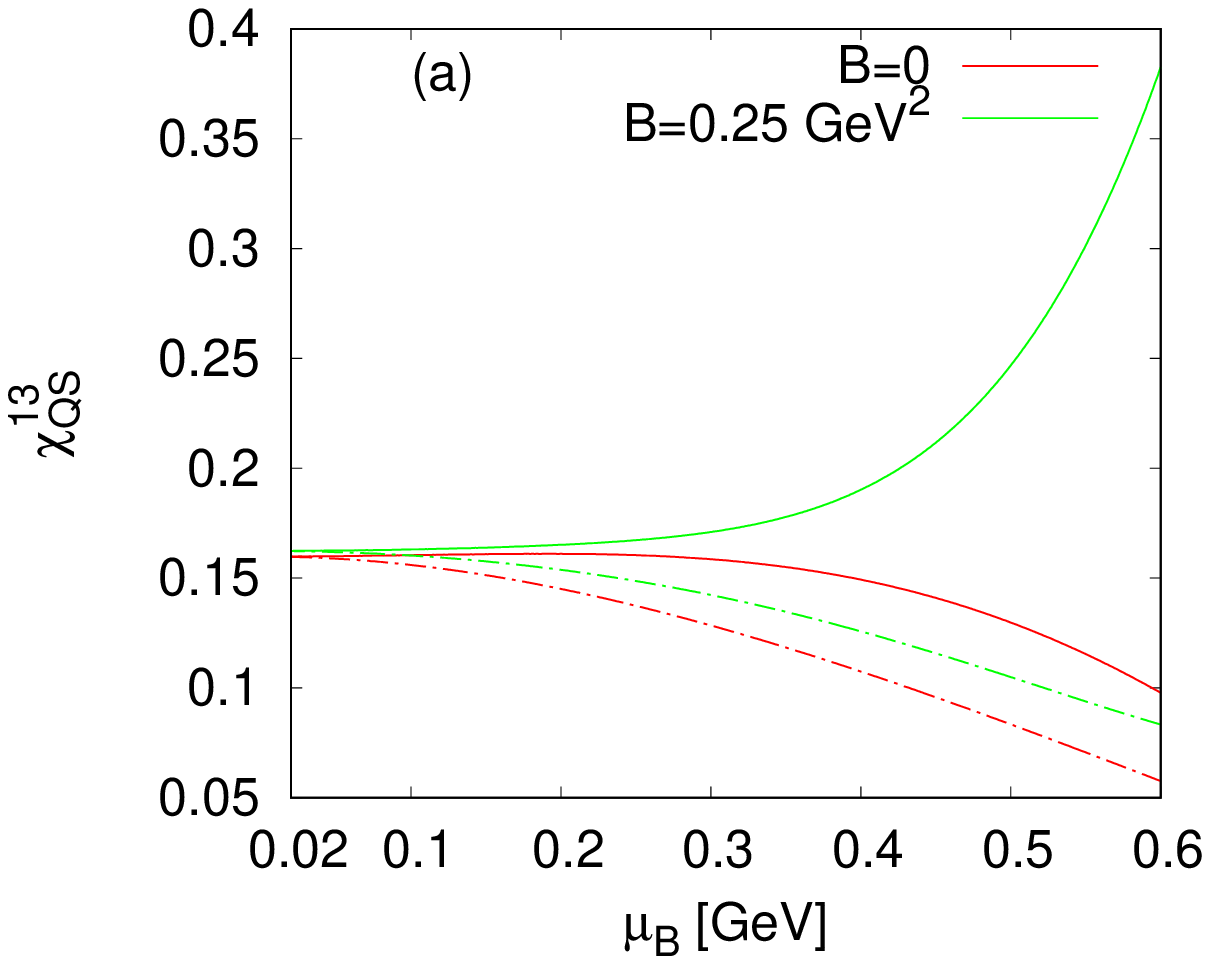}}
\subfloat{\includegraphics[scale=0.45]{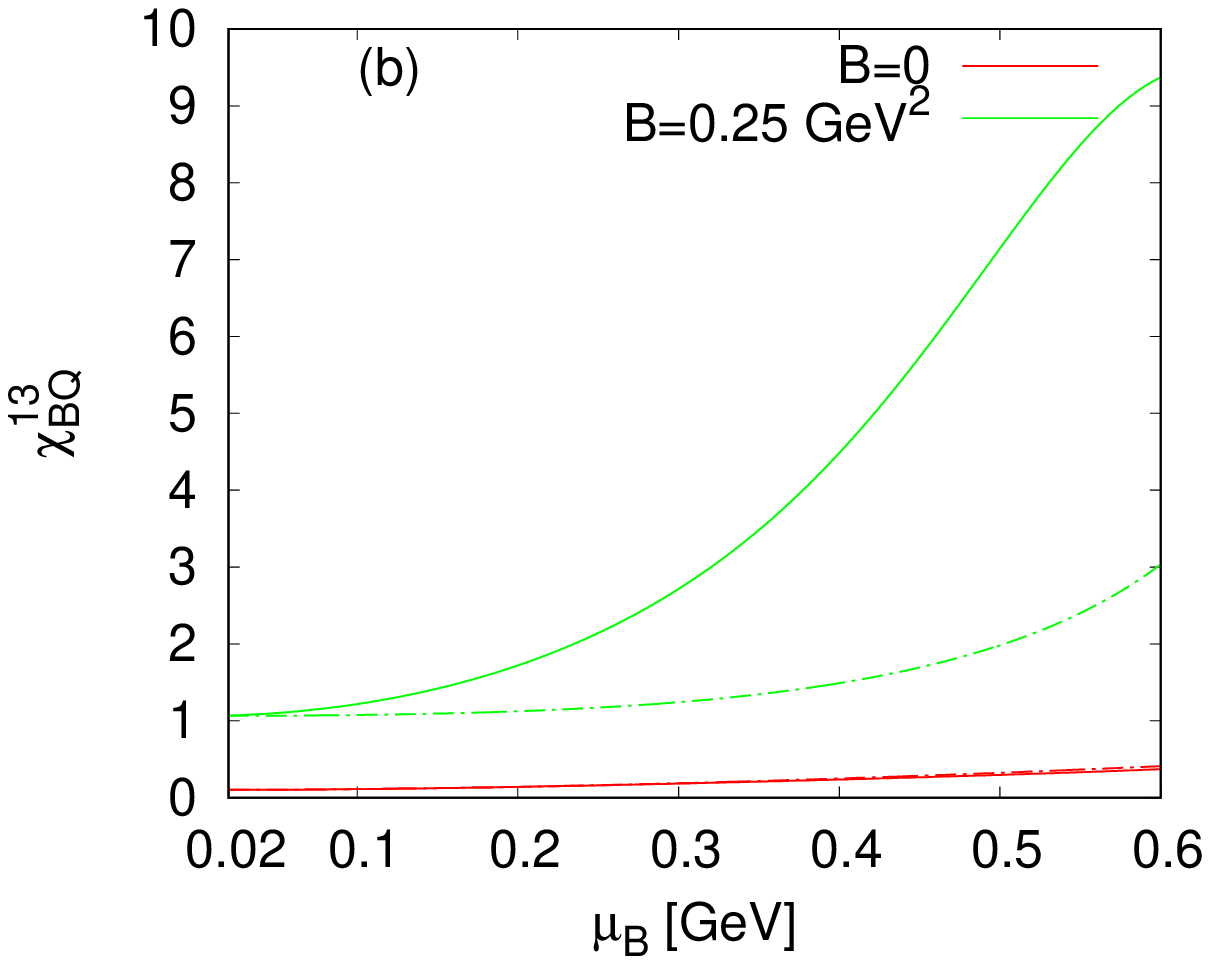}}
\subfloat{\includegraphics[scale=0.45]{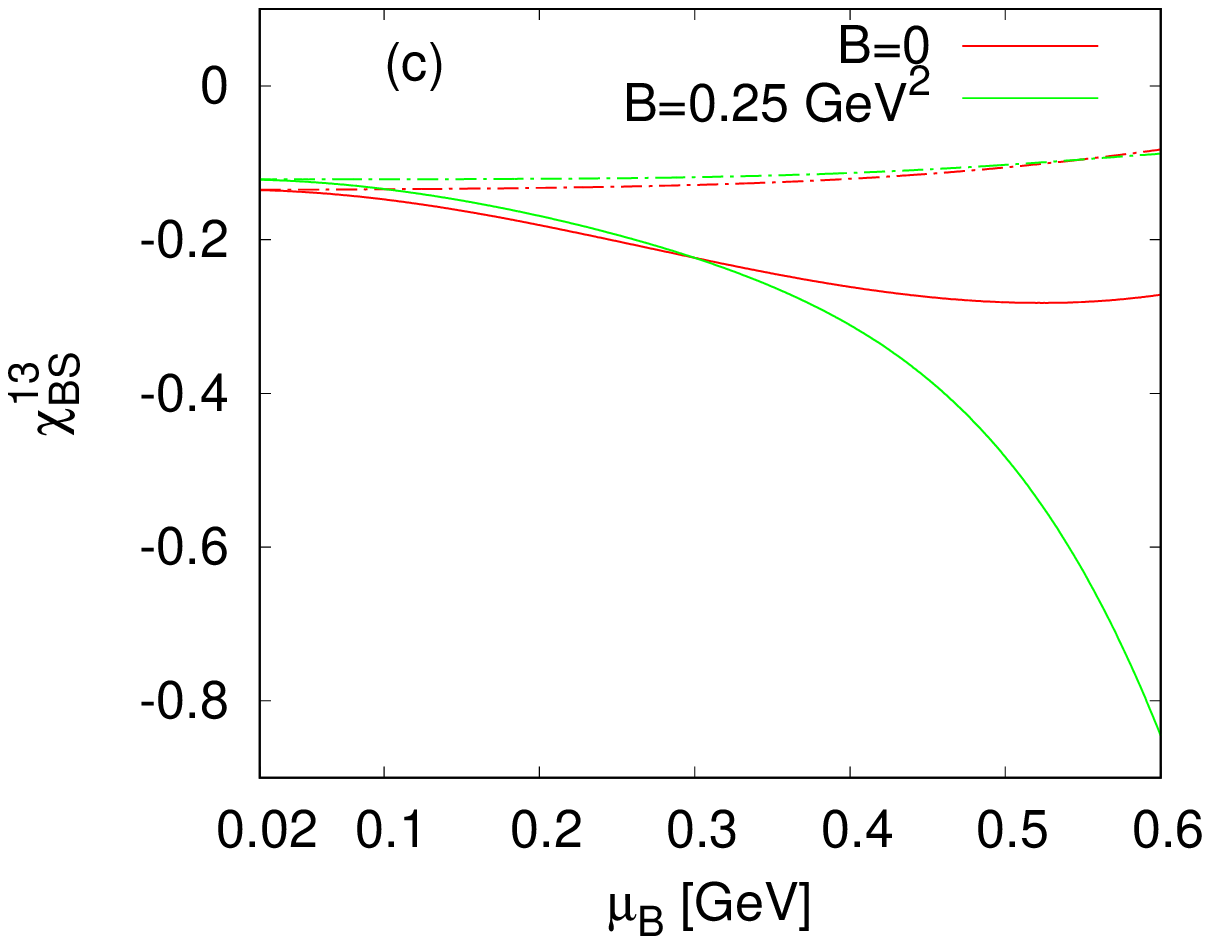}}
\caption{}{ $\chi^{13}_{QS}$, $\chi^{13}_{BQ}$ and $\chi^{13}_{BS}$ along the
freeze-out curve with and without B. Here the solid line
corresponds to the conserved charge correlations without charge conservation and the
dash-dotted line corresponds to the conserved charge correlations with charge conservation.}
\label{Fig.9}
\end{figure}
 
The higher order mixed correlations are presented in Fig.8 and Fig.9 respectively.
The charge conservation always diminishes the absolute value of the correlations.
The higher order mixed correlations (absolute value) are always larger compared to
the lower order mixed correlations. 

\subsection{Beam energy dependence of the products of moments}

Different moments such as mean (M), standard deviation ($\sigma$),
skewness (S) and kurtosis (k) of the conserved charges are measured 
experimentally to characterize the shape of the charge distributions.
The products of moments are related to the susceptibilities by
the following relation

\begin{equation} 
\frac{{\sigma}^2}{M}=\frac{{\chi}^2}{{\chi}^1},\hspace{10pt}
S{\sigma}=\frac{{\chi}^3}{{\chi}^2},\hspace{10pt}
k{{\sigma}^2}=\frac{{\chi}^4}{{\chi}^2}
\end{equation}
 
Fig.10 shows the products of different moments for net-proton, net-charge and net-kaon
as a function of center of mass energy $\sqrt{s}$. Here the net-proton and the net-kaon acts 
as a proxy for the conserved charges of the net baryon and the net strangeness respectively. 
I also have compared the moments of the net-proton and the net charge with the available 
experimental data \cite{star1,star2}. Here, three different values 
of the magnetic field have been considered which is achieved at different center of 
mass collision energies.

\begin{figure}
\centering
\subfloat{\includegraphics[scale=0.45]{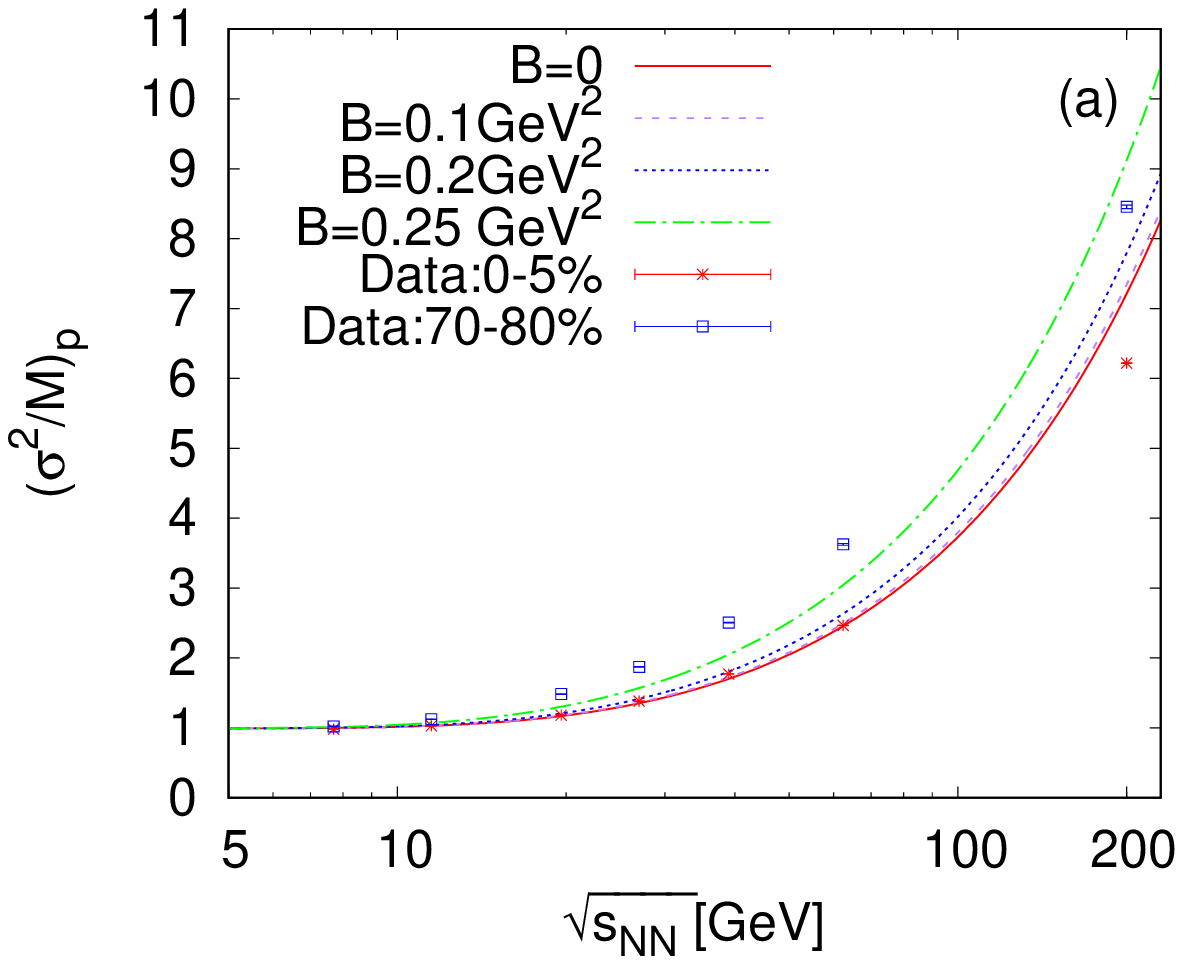}}
\subfloat{\includegraphics[scale=0.45]{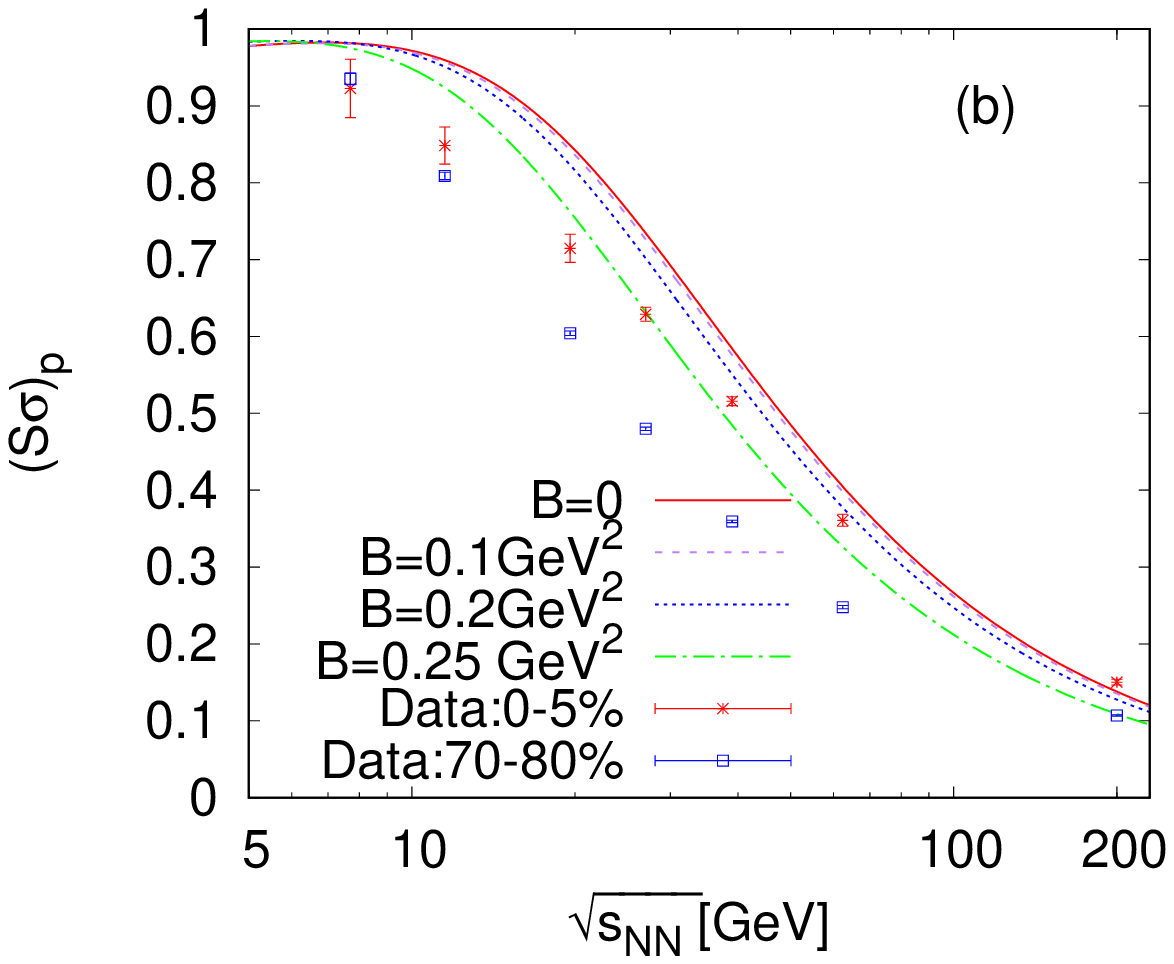}}
\subfloat{\includegraphics[scale=0.45]{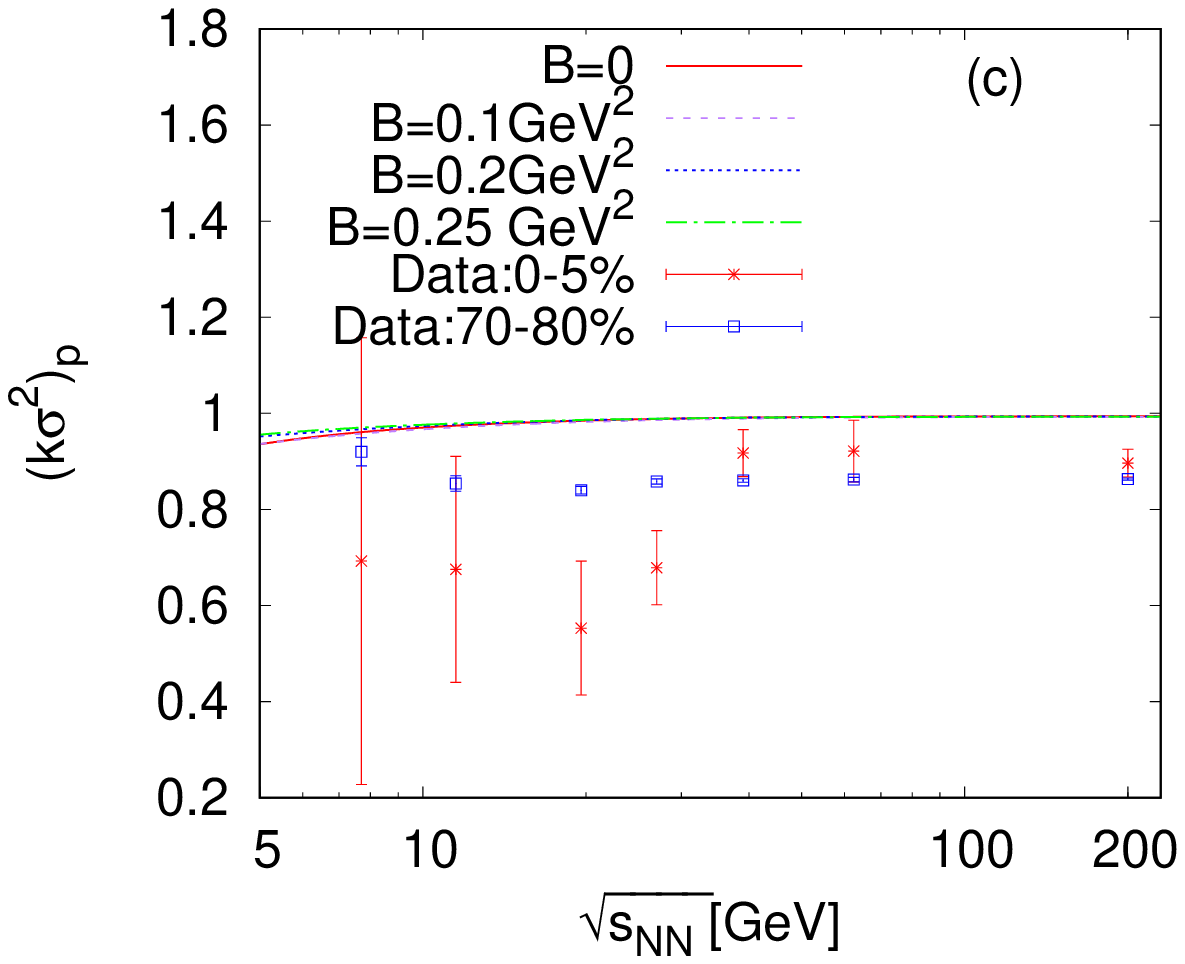}}\\
\subfloat{\includegraphics[scale=0.45]{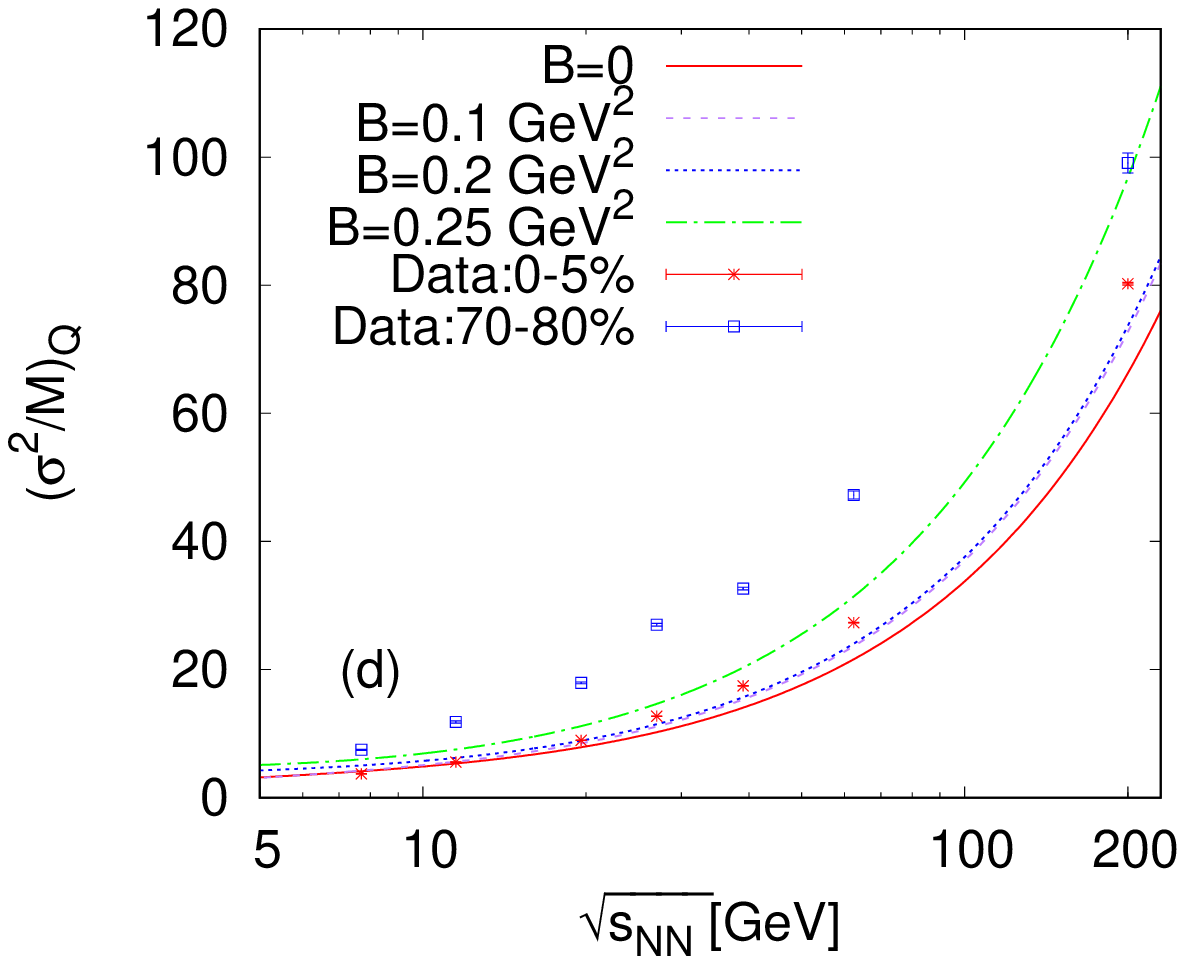}}
\subfloat{\includegraphics[scale=0.45]{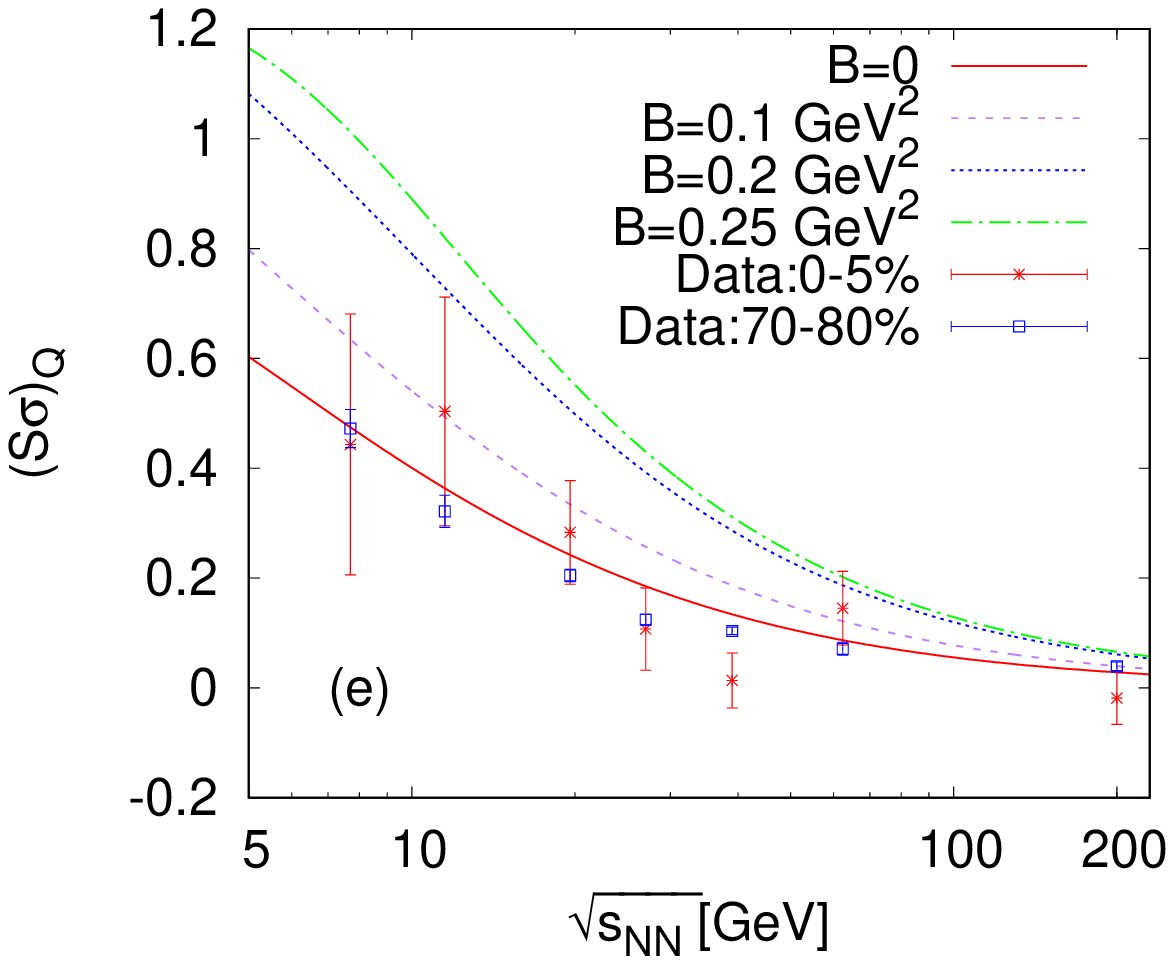}}
\subfloat{\includegraphics[scale=0.45]{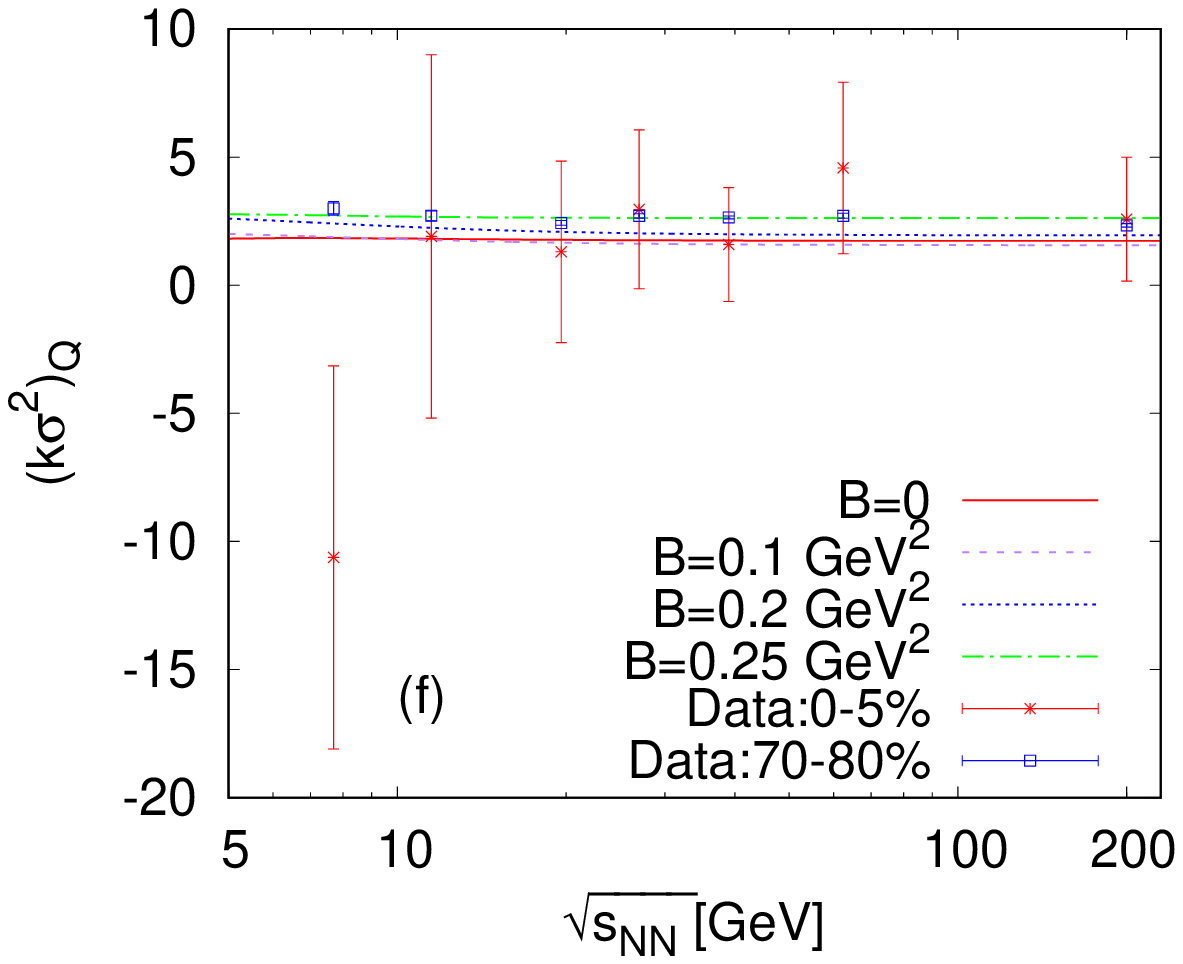}}\\
\subfloat{\includegraphics[scale=0.45]{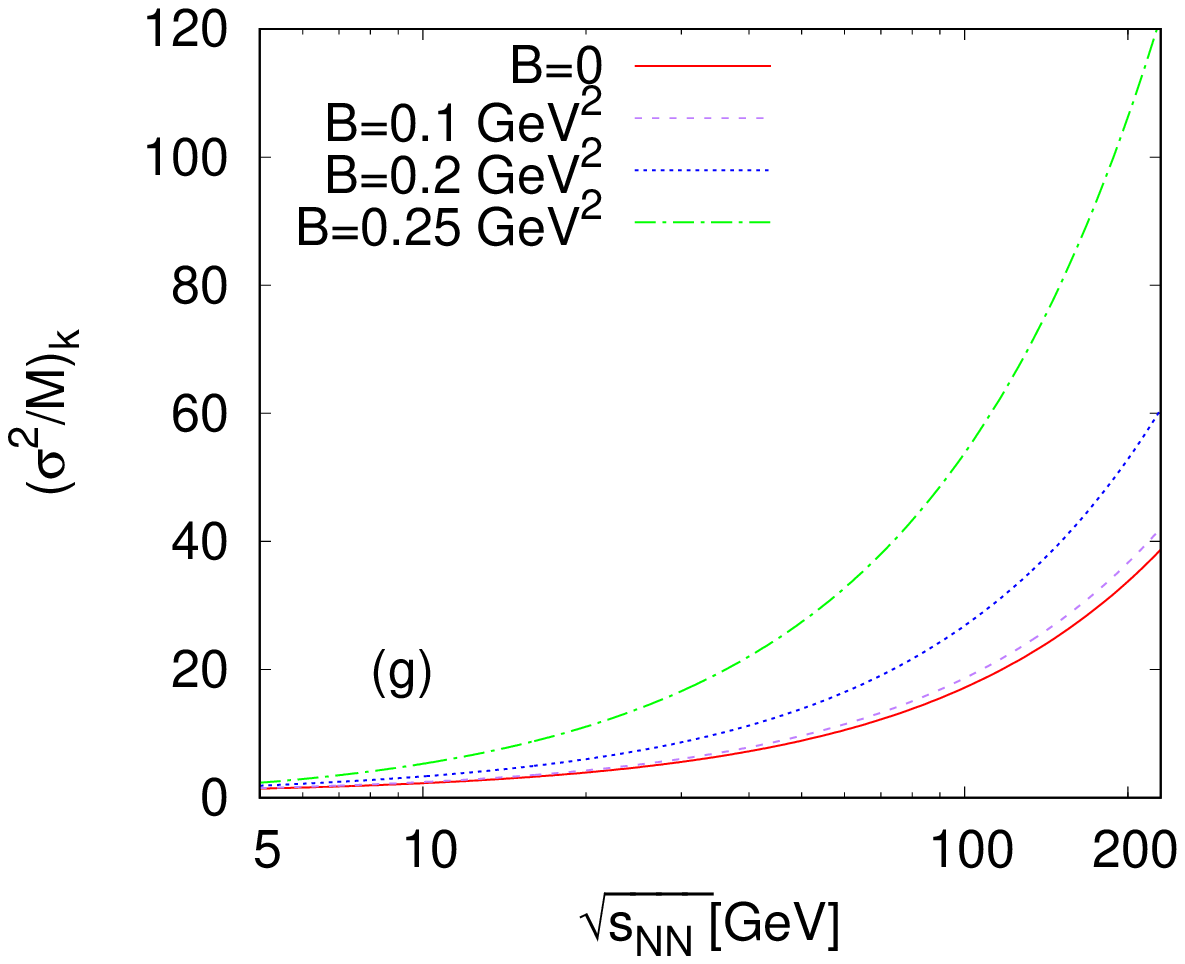}}
\subfloat{\includegraphics[scale=0.45]{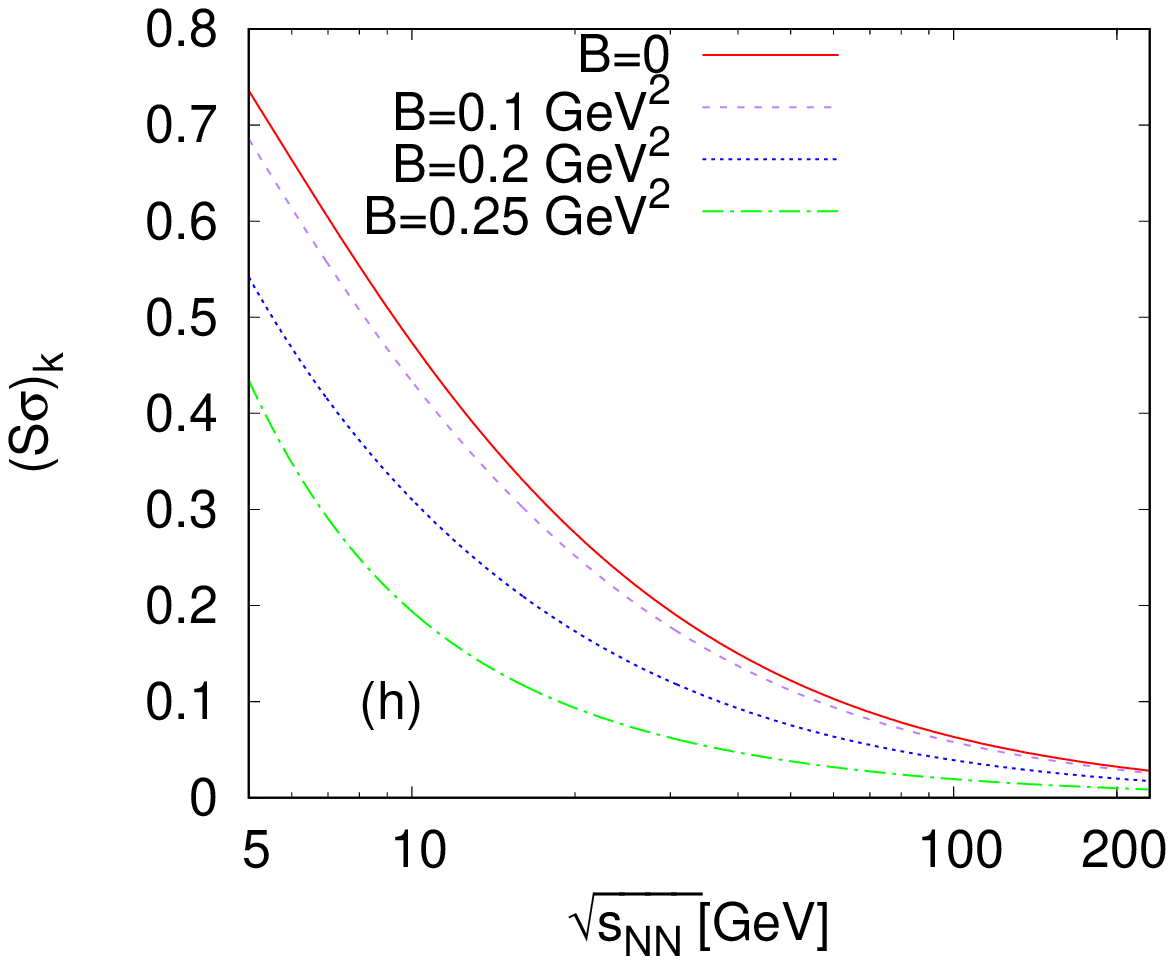}}
\subfloat{\includegraphics[scale=0.45]{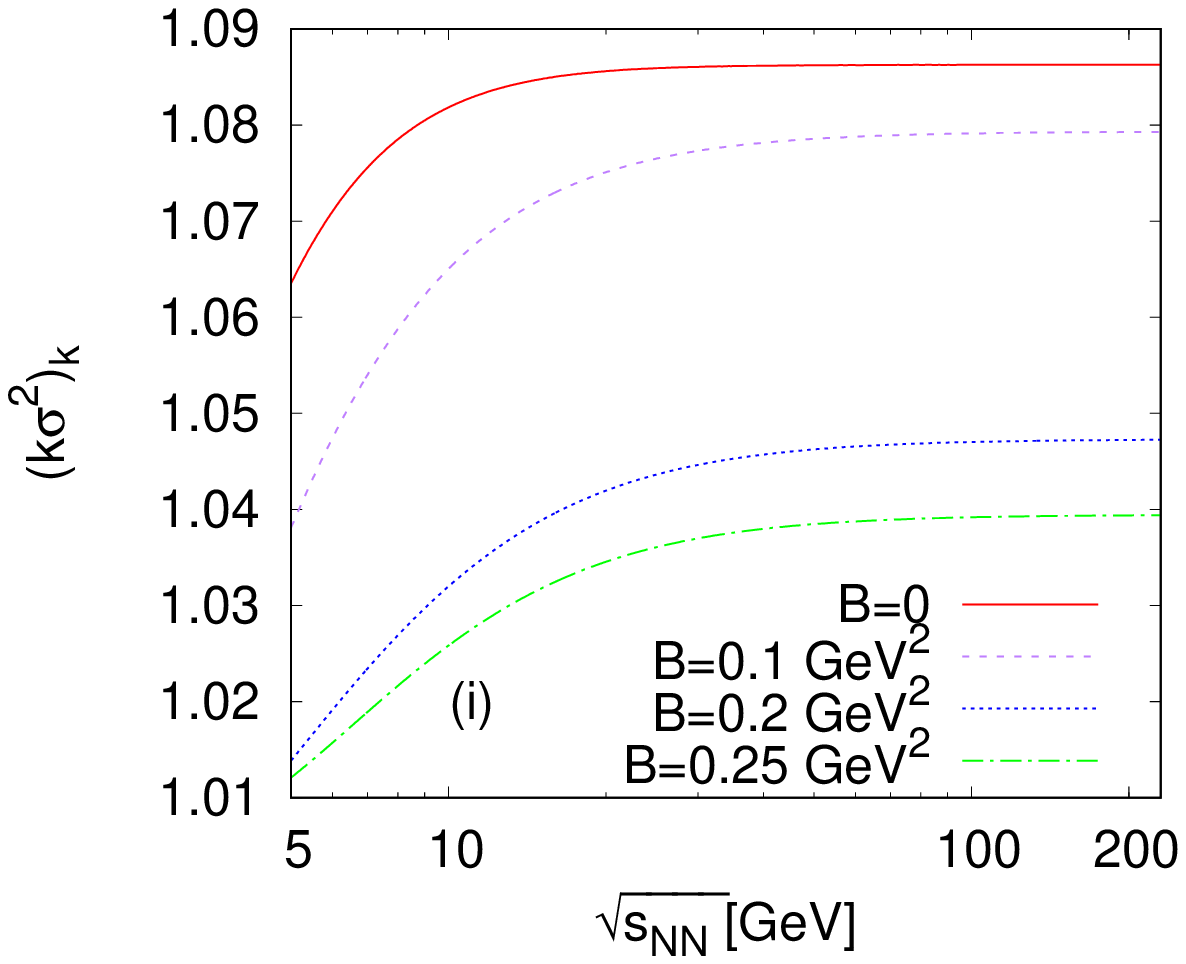}}
\caption{Products of moments of net-proton, net-charge and net-kaon as a function
of center of mass energy}
\end{figure}

These quantities have been measured at nonzero magnetic field in \cite{rajarshi} with fixed 
zero B freeze-out parameters. In ref.\cite{rajarshi}, these quantities have 
been measured in HRG model 
at nonzero magnetic field along the freeze-out curve where the freeze-out parameters 
are obtained from Eq.3 and Eq.4 at zero B. Here, one can see the  
results are very different with the IMC effect. The parameterization used in Eq.3 and Eq.4
to convert the freeze-out chemical potential and the temperature to center of
mass collision energy only exists for central collisions and zero magnetic field.
Since there is no parameterization to convert $\mu_{B}$ to center of mass energy $\sqrt{s}$ for
nonzero magnetic field and non-central collisions, I have used the same 
parameterization discussed in Eq.4 
to convert $\mu_{B}$ to the center of mass energy $\sqrt{s}$. However, I have used 
the temperature, the strange and the electric charge chemical potential corresponding to $\mu_{B}$ 
obtained with charge conservation. So, with this kind of parameterization, a one-to-one 
comparison with the experimental data is not expected to be perfect as the system
at different centralities and magnetic fields have different temperature, chemical potential 
and volume.

$(\sigma^2/M)_{p}$ increases with the increase in $\sqrt{s}$ for zero and nonzero B.
It increases slightly with increase in magnetic field at a particular 
center of mass energy. But, it decreases with magnetic field at at a particular
center of mass energy with fitted freeze-out parameters \cite{rajarshi}. $(S\sigma)_{p}$ increases
with increase in magnetic field at a particular collision energy with the IMC effect.
But, it decreases with the increase in magnetic field at a particular collision 
energy with the fitted freeze-out parameters \cite{rajarshi}. $(k{\sigma}^2)_{p}$ is of the
order 1 because net proton (net baryon) distribution is a skellam one. It does not depend on 
the magnetic field.

$(\sigma^2/M)_{Q}$ increases with respect to the center of mass energy. At a fixed energy, 
$(\sigma^2/M)_{Q}$ increases slowly with the increase in the magnetic field. This is in contrast 
to the results obtained from the fitted parameters where
this value decreases when the magnetic field increases \cite{rajarshi}. $(S\sigma)_{Q}$ increases
with the increase in the magnetic field. $(k{\sigma}^2)_{Q}$ increases slowly with the 
magnetic field. I have compared these quantities with data 0-5\% and 70-80\% centrality 
class. A one to one correspondence is not possible under this study, because for this one 
has to calculate the amount of magnetic field produced at different center of mass 
energies and its centrality dependence. The magnetic field increases with center 
of mass collision energy, so its effect on different products of moments are usually 
large at large $\sqrt{s}$.  

The most visualizing effect of the IMC appears in the net-kaon sector. $(\sigma^2/M)_{k}$ increases
with the increase in the magnetic field at a particular center of mass energy. However, it does 
not depend on the magnetic field when it is calculated from the fitted parameters 
\cite{rajarshi}. $(S\sigma)_{k}$ decreases with the increase in the
magnetic field. But it does not depend on the magnetic field when it is calculated from 
the fitted parameters \cite{rajarshi}. This is because kaon being a spin zero particle, 
its production is not enhanced significantly at nonzero B. So, when 
one uses fitted parameters ( same freeze-out temperature) to calculate the ratios at 
different magnetic fields, the ratios
are almost the same ( the ratio between two successive susceptibilities). However, when one 
uses the IMC effect i.e lower freeze-out temperature at nonzero B
compared to zero B, these ratios become different for different magnetic field. 
$(k{\sigma}^2)_{k}$ decreases with the increase in magnetic field.
This is same as the results obtained from the fitted parameters.

\section{CONCLUSIONS}

I have studied the IMC effect on the conserved charge fluctuations and the correlations 
along the freeze-out curve in the HRG model. The fluctuations 
and the correlations have been obtained with and without charge conservation.
At $B = 0$, the charge conservation does not play a role in the fluctuations 
(2nd and higher order) along the freeze-out curve for the conserved charges 
of electric charge and baryon number. But the charge conservation plays an important role
for the strange charge at $B = 0$. The charge conservation diminishes the fluctuations 
in the strange charge at $B = 0$ compared to the fluctuations without charge 
conservation. For nonzero B, the charge conservation plays a very important role.
If there is no charge conservation at nonzero B, then the fluctuations increase
by a huge amount compared to zero B at higher $\mu_{B}$. This is because
at higher $\mu_{B}$, there are more protons and charged baryons production at nonzero B.
The charge conservation diminishes the fluctuations along the freeze-out curve at nonzero B. 
I have compared these results taking zero B fitted parameters at nonzero B
and this is very different from the results obtained for nonzero B with and without
charge conservation. 

The correlations between different conserved charges also depend whether the charge 
conservation is taken into account or not. The correlations also should decrease
when charge conservation is taken because different charges production are restricted 
due to this. At $B = 0$, the correlation between electric charge and baryonic charge is almost 
the same with and without charge conservation. But this is not true for the other charge 
correlations like baryon and strangeness or electric charge and strangeness. The
correlation (modulus value) between the baryon and the strangeness increases without charge 
conservation compared to charge conservation at zero B. However, the
correlation between the electric charge and the strangeness decreases without charge
conservation compared to charge conservation at zero B. At nonzero B,
the correlation increases at higher $\mu_{B}$ without charge conservation. 
But when the charge conservation is taken into account, this decreases.
The correlation between the electric charge and the strangeness decreases without charge
conservation compared to charge conservation at nonzero B. However, at higher $\mu_{B}$,
this correlation increases at nonzero B.

I also have obtained the products of moments for net-proton, net-charge and 
net-kaon for three different magnetic fields and compared with the available 
experimental data. Here one can see the results are different from the zero B fitted 
parameters at nonzero B, this is clearly seen in net-kaon moments.
Here one assumes hadrons as point particles, but considering finite size of hadrons as
in an excluded volume model, these results will be almost invariant. This is because 
the freeze-out parameters are obtained from the universal freeze-out condition  
$E/N=\epsilon/n\simeq$ 1 GeV and the energy and the number of particles are extensive 
quantities, and the ratio is not affected by the excluded volume corrections.

\acknowledgments

I would like to thank Ajay K Dash and Rajarshi Ray for very useful 
discussions and suggestions.


\end{document}